%% file: ms.tex
\documentclass[apj]{emulateapj}

\newcommand{\eso}{ESO~1327--2041} 
\newcommand{\pks}{PKS~1327--206}
\newcommand{\hst}{{\sl HST}}

\newcommand{\Ha}{\ensuremath{{\rm H}\alpha}}
\newcommand{\Hb}{\ensuremath{{\rm H}\beta}}
\newcommand{\Hg}{\ensuremath{{\rm H}\gamma}}
\newcommand{\Hd}{\ensuremath{{\rm H}\delta}}
\newcommand{\He}{\ensuremath{{\rm H}\epsilon}}
\newcommand{\Hz}{\ensuremath{{\rm H}\zeta}}
\newcommand{\Heta}{\ensuremath{{\rm H}\eta}}
\newcommand{\Ht}{\ensuremath{{\rm H}\theta}}
\newcommand{\Hi}{\ensuremath{{\rm H}\iota}}
\newcommand{\Hk}{\ensuremath{{\rm H}\kappa}}

\newcommand{\CaII}{\ion{Ca}{2}}
\newcommand{\CIV}{\ion{C}{4}}
\newcommand{\HI}{\ion{H}{1}}
\newcommand{\HII}{\ion{H}{2}}
\newcommand{\MgII}{\ion{Mg}{2}}
\newcommand{\MnII}{\ion{Mn}{2}}
\newcommand{\NaI}{\ion{Na}{1}}
\newcommand{\NII}{[\ion{N}{2}]}

\newcommand{\eqw}{\ensuremath{{\mathcal W}_{\lambda}}}
\newcommand{\h}{\,\ensuremath{h_{70}^{-1}}}
\newcommand{\kms}{\ensuremath{{\rm km\,s}^{-1}}}
\newcommand{\lya}{\ensuremath{{\rm Ly}\alpha}}

\begin{document}

\title{The Quasar\,/ Galaxy Pair \pks\,/ \eso: \\ Absorption Associated with a Recent Galaxy Merger}
\author{Brian A. Keeney, John T. Stocke, Charles W. Danforth}
\affil{Center for Astrophysics and Space Astronomy, Department of Astrophysical and Planetary Sciences, University of Colorado, 389 UCB, Boulder, CO 80309; brian.keeney@colorado.edu}
\and
\author{Christopher L. Carilli}
\affil{National Radio Astronomy Observatory, P.O. Box O, Socorro, NM 87801}

\shorttitle{Quasar Absorption from a Recent Galaxy Merger}
\shortauthors{Keeney et~al.}

\begin{abstract}
We present \hst/WFPC2 broadband and ground-based \Ha\ images, \HI\ 21-cm emission maps, and low-resolution optical spectra of the nearby galaxy \eso, which is located $38\arcsec$ (14\h~kpc in projection) west of the quasar \pks.  Our \hst\ images reveal that \eso\ has a complex optical morphology, including an extended spiral arm that was previously classified as a polar ring.  Our optical spectra show \Ha\ emission from several \HII\ regions in this arm located $\sim5\arcsec$ from the quasar position ($\sim\,$2\h~kpc in projection) and our ground-based \Ha\ images reveal the presence of several additional \HII\ regions in an inclined disk near the galaxy's center.  Absorption associated with \eso\ is found in \HI\ 21-cm, optical, and near-UV spectra of \pks.  We find two absorption components at $cz_{\rm abs} = 5255$ and 5510~\kms\ in the \HI\ 21-cm absorption spectrum, which match the velocities of previously discovered metal-line components.  We attribute the 5510~\kms\ absorber to disk gas in the extended spiral arm and the 5255~\kms\ absorber to high-velocity gas that has been tidally stripped from the disk of \eso.  The complexity of the galaxy/absorber relationships for these very nearby \HI\ 21-cm absorbers suggests that the standard view of high redshift damped \lya\ absorbers is oversimplified in many cases.
\end{abstract}

\keywords{galaxies: interactions --- intergalactic medium --- quasars: absorption lines}

\section{Introduction}
\label{intro}

\HI\ absorption lines in the spectra of background quasars are one of the rare astrophysical phenomena to be discovered first at great distances, where the far-UV \lya\ transition redshifts into the optical.  Only with the advent of the {\sl Hubble Space Telescope} (\hst) and its UV spectrographs have we been able to study these absorption lines at low redshift and probe their associations with gas in and around nearby galaxies \citep*[e.g.,][]{morris93,bowen01,penton02,bowen02,jenkins03,tripp05,lehner09}.

\lya\ produces a ``forest'' of highly-ionized absorption at $z>1.8$ in ground-based spectra, as well as the occasional ($d{\mathcal N}/dz \sim 1$) high column density ($17 < \log{N_{\rm H\,I}} < 20.3~{\rm cm^{-2}}$) ``Lyman-limit systems'' \citep*[LLSs;][]{bergeron91,steidel92,prochaska99,prochaska10} and the rare ($d{\mathcal N}/dz \sim 0.1$) very high column density ($\log{N_{\rm H\,I}} \geq 20.3~{\rm cm^{-2}}$) ``damped \lya\ absorbers'' \citep*[DLAs;][]{wolfe05}.  Unlike the vast majority of lower column density absorbers, including the LLSs, the DLAs consist of primarily neutral gas and so are potentially more directly related to star formation.  Indeed, DLAs represent the largest neutral gas reservoir at each redshift of observation, which has been extended to $z > 5$ through the discovery of very high-$z$ QSOs and to $z < 1.8$ using \hst\ \citep{wolfe05}.  

There is no strong evidence for cosmological evolution in $\Omega_{\rm H\,I}$ over that entire range \citep{prochaska04}, although there is a factor $\sim\,$2 discrepancy between the lowest redshift DLA $\Omega_{\rm H\,I}$ values and those derived from \HI\ 21-cm emission surveys at $z \sim 0$ \citep{zwaan05}.  Further, there are very few DLAs known at $z < 1.8$ because observing time on \hst\ is too restrictive to allow a blind survey for low-$z$ DLAs.  A hybrid method of surveying only those QSOs with strong \MgII\ (and \ion{Fe}{2}) absorption has shown some success \citep*{rao06,nestor05,nestor06}, but the factor of two discrepancy remains.

High-$z$ DLAs are often modeled as thick gas disks associated with the progenitors of massive spiral galaxies \citep[e.g.,][]{wolfe00}, but this is largely out of a desire for simplicity since the optical/IR emitting galaxy associated with the DLA is rarely detected, or even detectable, using current ground-based or space-based telescopes.  The galaxy population responsible for DLAs is spread over a large range of luminosities \citep[0.001--1~$L^*$;][]{zwaan05,rosenberg03}, so it is not surprising that it has been difficult to detect the galaxies responsible for high-$z$ DLA absorption nor that the spread in DLA metallicities is $\sim\,$2~dex at any given redshift.  The $\log{N_{\rm H\,I}} \geq 20.3~{\rm cm^{-2}}$ limit makes good sense physically since above this limit the gas must be primarily neutral; however, systems with \HI\ column densities just below this limit (sometimes called ``sub-DLAs'' or ``super Lyman limit systems'') may share many of the same associated galaxy properties with the DLAs \citep{tripp05,peroux10a,peroux10b}.  

The absence of direct galaxy detection for many DLAs and sub-DLAs is in marked contrast to the situation for LLSs, which are found to be associated with nearby (impact parameter $\leq\,$50~kpc), luminous ($L \geq 0.1\,L^*$), gas-rich galaxies \citep{bergeron91,steidel92,barton09,chen10a,chen10b}.  The high luminosity and proximity of LLS host galaxies to the quasar sight line have been critical to their successful identification \citep[e.g.,][]{steidel95,churchill07,kacprzak10}.  The current consensus model for LLSs is gas in the halo of luminous galaxies although the physical processes that give rise to this gas are unclear and can encompass outflowing, unbound winds \citep{martin99,heckman01,adelberger03,kacprzak10} as well as outflowing or infalling galactic ``fountain'' material \citep*{keeney05,richter09,stocke10,chen10b}.  Theoretical models also predict that QSO absorption lines can be produced in the vicinity of galaxies from infalling cold mode accretion clouds \citep{keres09} or disrupted satellite galaxies \citep{york86,kacprzak10}. 

Discovery surveys for new low-$z$ absorber/galaxy associations, particularly DLAs and sub-DLAs, are needed to better resolve the associated galaxy from the QSO, to clearly detect the lower luminosity examples of associated galaxies, and to study the associated galaxy properties using all available ground-based and space-based techniques.  To this end we initiated a modest-sized \hst\ imaging project to study the detailed absorber/galaxy relationship for three very low-$z$ ($cz \leq 10,000$~\kms) \HI\ 21-cm absorbers: 3C~232\,/ NGC~3067, \pks\,/ ESO~132722--2040.6 (hereafter \eso), and PKS~2020--370\,/ Klemola~31A.  While each of these systems was initially discovered using \HI\ 21-cm absorption (which is only sensitive to DLAs and sub-DLAs) against a background radio-loud quasar, two of the three (the first two above) were also detected in \CaII~H \& K and \NaI~D absorption by ground-based telescopes \citep*{boksenberg78,stocke91,kunth84,bergeron87}. \citet[CvG92 hereafter]{carilli92} have used the Very Large Array (VLA) to map each of these systems in \HI\ 21-cm emission for comparison to the absorption line data. 

In a previous paper \citep{stocke10} we presented our new \hst\ imaging of the 3C~232/NGC~3067 system, which allowed us to conclude that the \HI\ 21-cm sub-DLA absorber in the spectrum of that quasar is very similar in its physical properties to high-velocity clouds (HVCs) found around the Milky Way \citep{keeney05,keres09,heitsch09,faucher-giguere10}.  We also found that this gas is infalling onto the disk of NGC~3067.  The simplicity of this QSO/galaxy pair --- relatively isolated galaxy viewed at an intermediate inclination angle --- allowed for an easy interpretation of the cloud kinematics and properties.

In this paper we present our \hst\ imaging results for the second pair, \pks\,/ \eso, a much more complex galaxy system with polar ring \HI\ gas, which we show is likely due to a direct galaxy collision between a gas poor lenticular galaxy and a late-type spiral.  The quasar absorber consists of two velocity components separated by $\sim\,$250~\kms\ in both metal lines and \HI\ 21-cm absorption.  However, \lya\ itself in this system is not directly observable due to a high-$z$ LLS, so we do not know for certain whether this absorber complex is a DLA or a sub-DLA.  Regardless, this absorber/galaxy pair clearly shows the potential complexity of such systems.

We have obtained new VLA \HI\ 21-cm emission maps, a new \HI\ 21-cm absorption spectrum, ground-based \Ha\ imaging and long-slit spectroscopy of various galaxy components, as well as a new optical spectrum of the quasar at moderate ($\sim\,$2~\AA) resolution, in addition to the aforementioned \hst\ continuum imaging of this $\sim L^*$ galaxy merger at $z=0.0178$ ($cz \approx 5500$~\kms).  We also present an archival \hst/FOS spectrum of \pks\ which detects several near-UV metal lines including \ion{Mg}{1}, \MgII, \ion{Fe}{2}, and \MnII\ at the redshift of \eso. In \S\,\ref{gal} we present the new \hst\ and ground-based images, the \HI\ 21-cm emission maps, and the optical spectra of various regions in \eso. In \S\,\ref{qso} we show the \HI\ 21-cm, optical, and near-UV spectra of the quasar \pks\ that we have analyzed. In \S\,\ref{nucleus} we speculate on the nature of a nearly unresolved knot of starlight emission discovered in our \hst\ images which may be a hyper-compact galaxy nucleus ejected from \eso\ as a result of the galaxy collision now on-going. In \S\,\ref{conclusion} we summarize the results of our observational campaign on this system and discuss the importance of this system for interpreting high-$z$ DLAs and sub-DLAs.  We assume $h_{70} = H_0 / (70~{\rm km\,s^{-1}\,Mpc^{-1}})$ and a distance to \eso\ of $80\pm5$\h~Mpc throughout this paper.

\section{The Interacting Galaxy Pair \eso}
\label{gal}

Early CCD images of \eso\ revealed its disturbed optical morphology, which includes an elongated nucleus (oriented roughly north-south) with a central bulge, an east-west ``polar ring'' that extends from the galaxy nucleus to the quasar position, and a bright plume that extends $45\arcsec$ northeast of the galaxy nucleus \citep{giraud86}.  Subsequent \HI\ 21-cm emission maps of \eso\ confirmed the presence of a regularly-rotating gaseous ring extending in the direction of \pks\ (CvG92).  This complex morphology suggests that \eso\ is the result of a recent major merger and/or a significant tidal interaction and that any absorption lines probing this system will also be quite complex.

We have obtained high-resolution observations of \eso\ at multiple wavelengths, allowing us to study this galaxy in unprecedented detail.  These observations include:  \hst\ broadband (\S\,\ref{gal:hstimg}) and ground-based \Ha\ images (\S\,\ref{gal:apoimg}) of \eso, \HI\ 21-cm emission maps of \eso\ and \pks\ (\S\,\ref{gal:vla}), and long-slit optical spectra of several regions of \eso\ (\S\,\ref{gal:apospec}).

\subsection{Broadband \hst\ Images}
\label{gal:hstimg}

\begin{figure*}
\epsscale{1.00}
\centering \plottwo{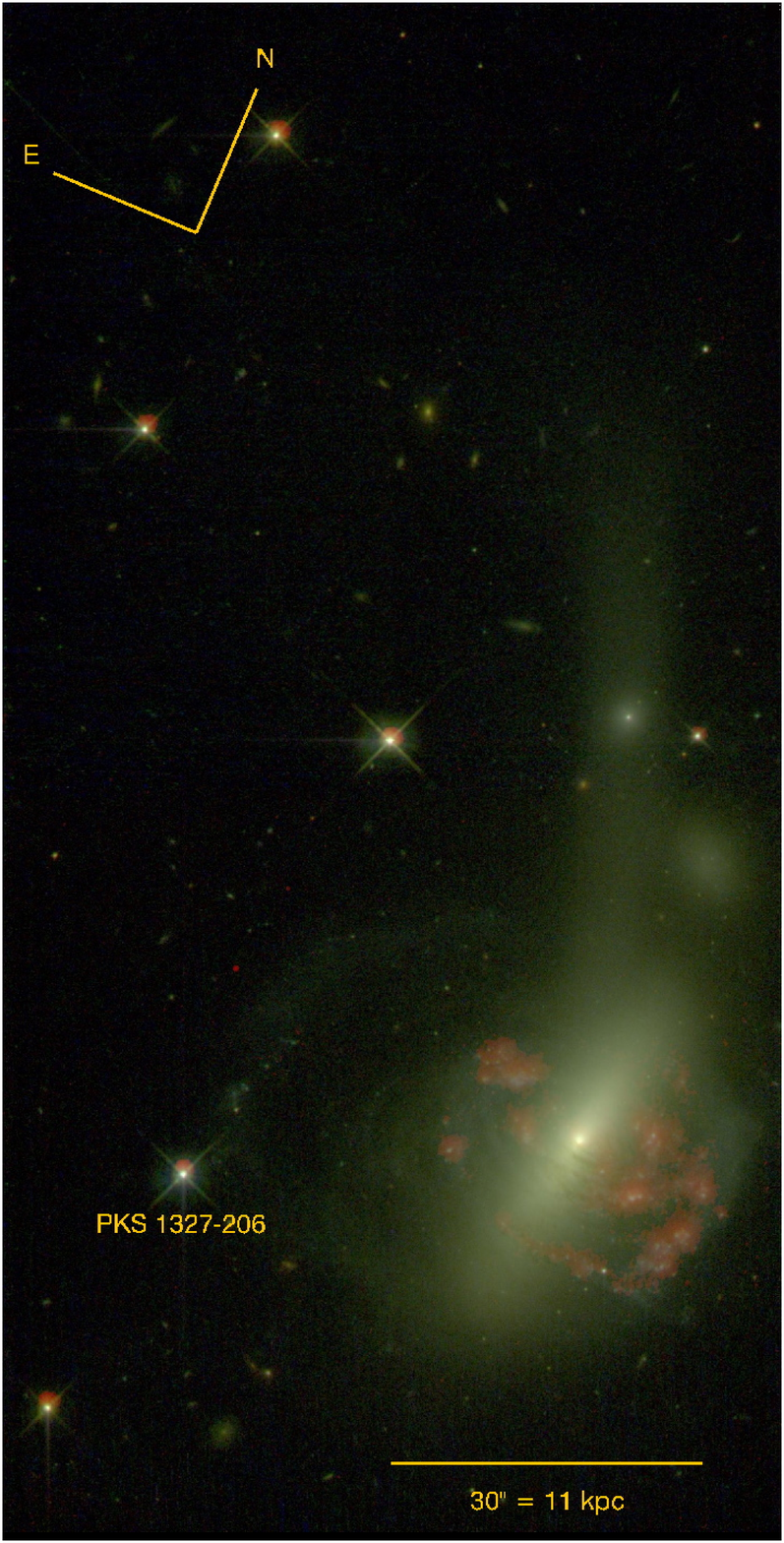}{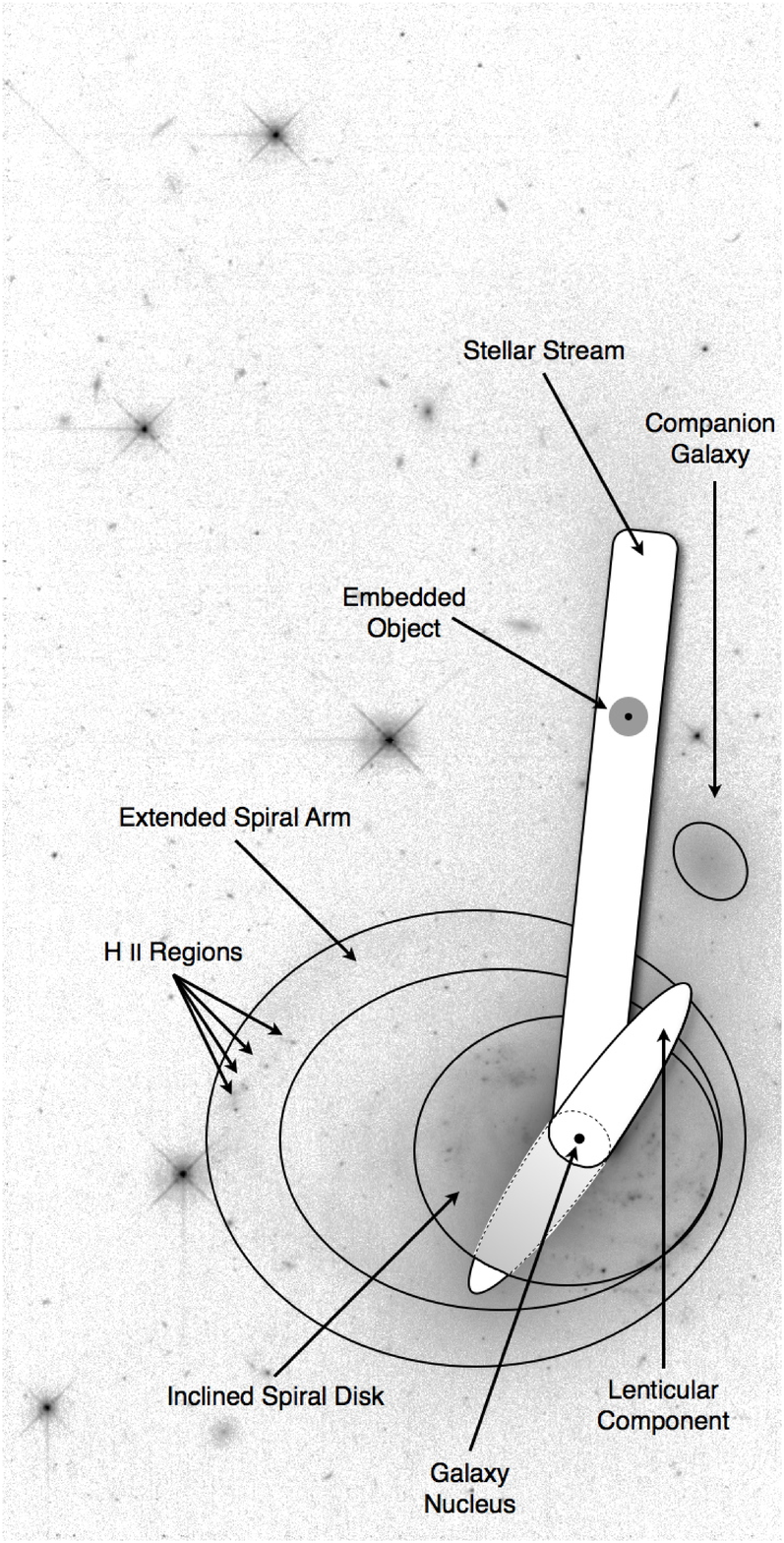}
\caption{{\em Left:}  Color composite \hst/WFPC2 image of \eso.  Data from the F450W filter is displayed in blue, F675W in green, and F791W in red.  The image orientation and the location of the nearby quasar \pks\ are labelled.  The lower resolution red regions near the center of the galaxy are \HII\ regions found in our ground-based \Ha\ images (\S\,\ref{gal:apoimg}).  The field-of-view of the image is approximately $75\arcsec \times 150\arcsec$, which corresponds to a physical scale of $\sim28\times56$\h~kpc at the galaxy redshift.  {\em Right:}  Schematic diagram of \eso\ overlaid on the F675W WFPC2 image.  All galaxy components discussed in \S\,\ref{gal} have been labelled.
\label{fig:hst}}
\end{figure*}

\eso\ was observed with the Wide-Field Planetary Camera 2 (WFPC2) on board \hst\ on 16~Apr~2007 for a total of 4200\,s in the F675W ($R$-band) filter and 2400\,s in each of the F450W (Wide $B$-band) and F791W (Wide $I$-band) filters as part of GO program 10925 (PI:  J. Stocke).  The images were reduced and coadded using the {\tt stsdas} package of IRAF, as described in the WFPC2 Data Handbook\footnote{The WFPC2 Data Handbook can be found at {\tt http://www.stsci.edu/hst/wfpc2/}.}.  The orientation of our observations were such that \eso\ and \pks\ fell only on WF2 and WF3, so we did not analyze the WF4 or PC chips of the detector.  Our reduced images reach limiting 3$\sigma$ Vega magnitudes of 26.9, 26.5, and 26.0 for point sources in the F450W, F675W, and F791W filters, respectively.

A color composite image of the WFPC2 data (WF2 and WF3 chips only) is shown in the left panel of Figure~\ref{fig:hst}, where the F450W data are displayed in blue, the F675W data in green, and the F791W data in red.  The lower-resolution red areas in the center of \eso\ are \HII\ regions detected in our ground-based \Ha\ images (\S\,\ref{gal:apoimg}).  The field-of-view of Figure~\ref{fig:hst} is approximately $75\arcsec \times 150\arcsec$, which corresponds to a physical scale of $\sim 28\times56$\h~kpc at the redshift of \eso.

All of the features identified by \citet{giraud86} and CvG92 are present in the WFPC2 images, but the superior spatial resolution of \hst\ reveals several new features that we have labelled in a schematic of the WFPC2 images in the right panel of Figure~\ref{fig:hst}.  In particular, the central region of \eso\ is resolved into two components that are blended together at ground-based resolution:  a lenticular component that corresponds to the ``elongated nucleus'' of \citet{giraud86}, and an inclined spiral that contains several \HII\ regions.  These components are clearly co-spatial since \HII\ regions in the spiral disk are seen in front of the starlight from the lenticular component to the south of the galaxy nucleus but are occulted by the lenticular starlight to the north of the nucleus (i.e., the inclined disk is seen both in front of and behind the lenticular component).  The ``plume'' of material extending NNE from the galaxy nucleus is resolved into a stellar stream with a compact embedded source that we speculate in \S\,\ref{nucleus} is the ejected nucleus of the inclined spiral.  There is also a small companion galaxy located just north of the stellar stream, approximately halfway between the embedded source and the lenticular nucleus.

The previously identified ``polar ring'' is now resolved to be the outermost arm of the inclined spiral in the central regions of \eso.  This arm may be in the process of being tidally stripped and passes quite close to the \pks\ sight line.  We have detected several \HII\ regions in this extended arm located $\sim5\arcsec$ ($\sim\,$2\h~kpc) north of \pks\ (see Fig.~\ref{fig:hst} and \S\,\ref{gal:apospec}).  

CvG92 found that the gas in the ``polar ring'' of \eso\ is rotating regularly with lower velocities ($\sim\,$5250~\kms) in the west and higher velocities ($\sim\,$5500~\kms) in the east.  This rotation is not surprising since we now identify this feature with an extended spiral arm.  However, whether we call the gas that extends from the galaxy nucleus in the direction of \pks\ a ``polar ring'' or an ``extended spiral arm'' is largely a matter of semantics since the standard picture of polar-ring galaxies is that they are formed via galaxy mergers or the accretion of a companion galaxy or IGM filament \citep{moiseev09,bournaud03}.

\subsection{H$\alpha$ Images}
\label{gal:apoimg}

Ideally, we would have obtained \Ha\ images of \eso\ with \hst\ as well, but neither WFPC2 nor ACS had a narrowband filter suitable for observing \Ha\ at the redshift of \eso.  Instead the galaxy was observed at Apache Point Observatory (APO) on 24~Mar~2006 using the SPIcam imager of the ARC 3.5-m telescope.  Redshifted \Ha\ + \NII\ images were obtained in $1\farcs3$ seeing for a total of 3000\,s using a filter with a central wavelength of 6650~\AA\ and ${\rm FWHM} = 80$~\AA. Additionally, \eso\ was observed for 1800\,s in $1\farcs4$ seeing through an off-band filter with ${\rm FWHM} = 100$~\AA\ and a central wavelength of 6450~\AA\ in order to measure the strength of the stellar continuum.  All images were reduced and coadded using standard IRAF procedures.  The off-band image was then scaled so that its sky level matched that of the \Ha\ on-band image before continuum subtraction.  The resulting image should only contain flux from the \Ha\ and \NII\ emission lines, but there is some residual flux in the cores of bright stars due to the slight mismatch in seeing between the on-band and off-band images.  Our final continuum-subtracted \Ha\ image has been overlaid in red on the WFPC2 data in Figure~\ref{fig:hst}.  Only pixels with $F(\Ha) > 10^{-17}~{\rm ergs\,s^{-1}\,cm^{-2}}$ are shown, corresponding to areal star formation rates $>0.005\,h_{70}^{-2}~{\rm M_{\Sun}\,yr^{-1}\,kpc^{-2}}$ at the distance of \eso\ \citep{kennicutt98}.  

Figure~\ref{fig:hst} clearly highlights several \HII\ regions in the inclined spiral disk of \eso.  These \HII\ regions have an integrated \Ha\ flux of $F(\Ha) = (7.2\pm1.4)\times10^{-14}~{\rm ergs\,s^{-1}\,cm^{-2}}$ assuming a \NII/\Ha\ ratio of $0.3\pm0.2$ \citep[derived as in][]{kennicutt08}. At our assumed distance to \eso\ of $80\pm5$\h~Mpc, this corresponds to a luminosity of $(5.5\pm1.3)\times10^{40}\,h_{70}^{-2}~{\rm ergs\,s^{-1}}$ and a star formation rate of $0.44\pm0.10\,h_{70}^{-2}~{\rm M_{\Sun}\,yr^{-1}}$ \citep{kennicutt98}.  So, despite the morphological evidence for a significant merger event in the recent past of \eso, this interaction is not currently driving a major starburst episode.

The semi-stellar knots just NNW of \pks\ (labelled as \HII\ regions in the right panel of Fig.~\ref{fig:hst}) show no obvious \Ha\ emission in our narrowband images.  Nevertheless, a deep long-slit optical spectrum clearly detects very weak \Ha\ emission associated with these knots (see \S\,\ref{gal:apospec}), solidifying our earlier identification.

\subsection{\HI\ 21-cm Emission Maps}
\label{gal:vla}

\begin{figure}
\epsscale{1.00}
\centering \plotone{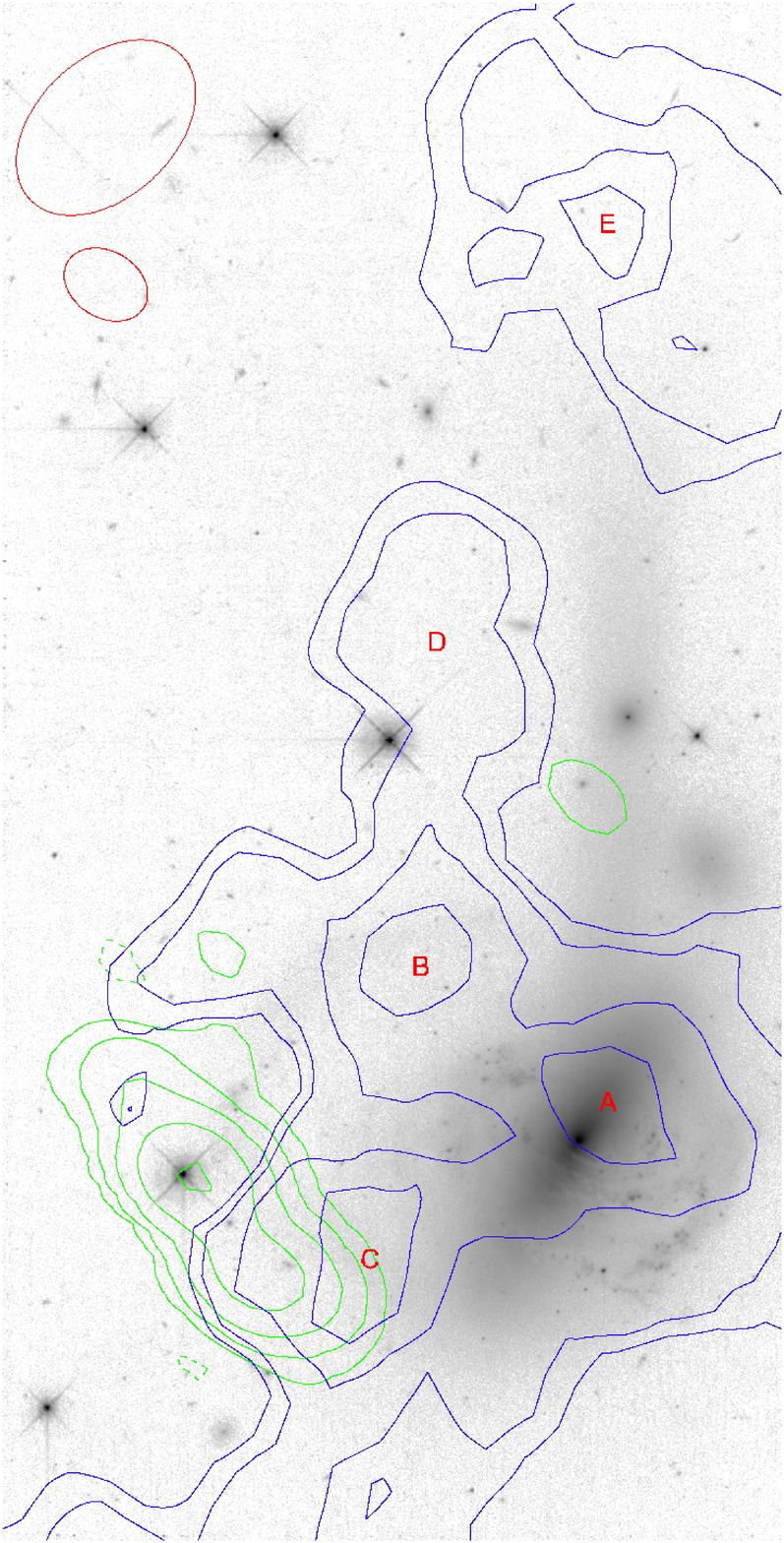}
\caption{Natural-weighted \HI\ 21-cm emission contours for \eso\ ({\em blue}) and uniform-weighted L-band continuum contours for \pks\ ({\em green}) overlaid on our F675W WFPC2 data. The \HI\ contour levels are 2.5, 5, 10, and $15\times10^{20}~{\rm atoms\,cm^{-2}}$ and the continuum contour levels are -3, 3, 9, 27, 81, and 243~${\rm mJy\,beam^{-1}}$.  The large red ellipse ($20\arcsec \times 13\arcsec$) in the upper-left corner of the image is the restoring beam for the \HI\ contours; the smaller ellipse ($9\arcsec \times 6\arcsec$) is the restoring beam for the continuum contours, and the labelled regions (A--E) correspond to the regions in Table~\ref{tab:HI}.  
\label{fig:vlamap}}
\end{figure}

\eso\ and \pks\ were observed by the Very Large Array (VLA) in the CnB configuration for 18.74~hours between 18~Jun and 4~Jul~2005.  Data were obtained in L-band (1.4~GHz) centered at a heliocentric velocity\footnote{We have reduced the VLA data using the optical definition of velocity, $v/c = (\nu_0 - \nu)/\nu = (\lambda - \lambda_0)/\lambda_0 = z$, rather than the traditional radio definition.} of 5370~\kms\ using 63 spectral channels with a width of 10.4~\kms\ (48.8~kHz) per channel.  To improve sensitivity and resolution, these data were combined with the data of CvG92 after flagging and calibration.  The CvG92 data were obtained with the VLA C-array on 20~Dec~1990 with the same channel configuration described above.  The addition of the CvG92 data increases the total exposure time of our observations to 25.8~hours.  All reductions for both data sets were performed with the Common Astronomy Software Applications (CASA) Beta Release, Version 2.0.

Figure~\ref{fig:vlamap} shows \HI\ 21-cm (blue) and 1.4-GHz continuum (green) emission contours overlaid on the F675W WFPC2 data.  The large red ellipse in the upper-left corner of the image is the restoring beam of the natural-weighted \HI\ contours and the smaller ellipse is the restoring beam of the uniform-weighted continuum contours.  \pks\ has a double-lobed morphology and a peak continuum flux of $280~{\rm mJy\,beam^{-1}}$ at this resolution.  There is also a weak continuum source (peak flux $\sim 5~{\rm mJy\,beam^{-1}}$), first detected by CvG92, located $\sim\,$10\arcsec\ south of the embedded object in the stellar stream (see Fig.~\ref{fig:hst}) that Figure~\ref{fig:vlamap} reveals is associated with a background galaxy and not the embedded object itself.

Continuum sources were subtracted from the data cube before imaging the \HI\ emission  using the CASA task {\tt uvcontsub} to perform a channel-by-channel linear fit to the continuum.  We believe that the hole in the \HI\ distribution toward \pks\ is due to the \HI\ 21-cm absorption against the quasar background (see \S\,\ref{qso:vlaspec}).  The \HI\ contours in Figure~\ref{fig:vlamap} show three peaks:  one (labelled ``A'' in Fig.~\ref{fig:vlamap}) near the nucleus of the lenticular component of \eso\ and two (``B'' and ``C'') that are likely associated with the extended spiral arm.  There is also a region of more diffuse emission (``D'') that extends in the direction of the stellar stream but is offset approximately one beam width to the east and, finally, a large diffuse region (``E'') near the northern edge of the image.  

Table~\ref{tab:HI} lists the positions, velocities, and \HI\ masses of these regions.  We find that gas in the extended spiral arm is rotating regularly with lower velocities in the south and west (regions A \& C) and higher velocities in the north and east (region B).  The more diffuse extended regions of \HI\ 21-cm emission (regions D \& E) have velocities that are $\sim\,$150~\kms\ lower than the closest disk gas in region B.

\input tab1.tex

Figure~\ref{fig:vlamap} can be used to determine how the recent merger history of \eso\ has affected its 21-cm emission properties compared to more isolated galaxies.  The \HI-emitting regions of typical spiral galaxies extend about twice as far as the regions of optical emission \citep[e.g.,][]{broeils97,noordermeer05}, which suggests that the ``pre-merger'' \HI\ extent of \eso\ was $\sim\,$60\arcsec\ ($\sim\,$23\h~kpc).  The measured angular extent of regions A--C in Figure~\ref{fig:vlamap} is $\sim\,$60\arcsec, but increases to $\sim\,$80\arcsec\ ($\sim\,$31\h~kpc) if region D is included.  Thus, the size of the \HI-emitting region of \eso\ has increased only modestly as a result of its recent interaction.

However, the galaxy interaction {\em has} caused \eso\ to become noticably \HI-deficient as compared to more isolated galaxies. The total \HI\ mass of regions A--E of \eso\ is $M_{\rm H\,I} = (9.6\pm0.6)\times10^8~M_{\Sun}$ (see Table~\ref{tab:HI}), but late-type spirals with the same luminosity and optical size as \eso\ tend to have \HI\ masses of $M_{\rm H\,I} \sim 4\times10^9 M_{\Sun}$ \citep{haynes84}.  This deficiency can easily be explained if most of the \HI\ in \eso\ was ionized during the interaction.  Ionization considerations may also explain why the interaction failed to dramatically increase the \HI\ extent of \eso.

\subsection{Optical Spectra}
\label{gal:apospec}

\eso\ was observed with the DIS spectrograph of the ARC 3.5-m telescope at APO on 1~May~2008 and 27-28~Feb~2009 using a $1\farcs5$ slit and the B400+R300 gratings. These gratings cover the wavelength range 3600--9000~\AA\ at 6-7~\AA\ resolution except for a gap from $\sim\,$5300-5700~\AA\ caused by a dichroic in the spectrograph.  Data were obtained at three different slit positions:  the first slit position was oriented along the major axis of the lenticular component of \eso, the second position was along the stellar stream with the slit passing through the embedded compact object, and the third position connected the compact object embedded in the stellar stream and \pks\ with the slit also passing through the \HII\ regions in the extended spiral arm.  Observations of Feige~34 taken each night were used for flux calibration, and all data were reduced with a combination of standard IRAF tasks and custom IDL routines.  

\begin{figure}
\epsscale{1.00}
\centering \plotone{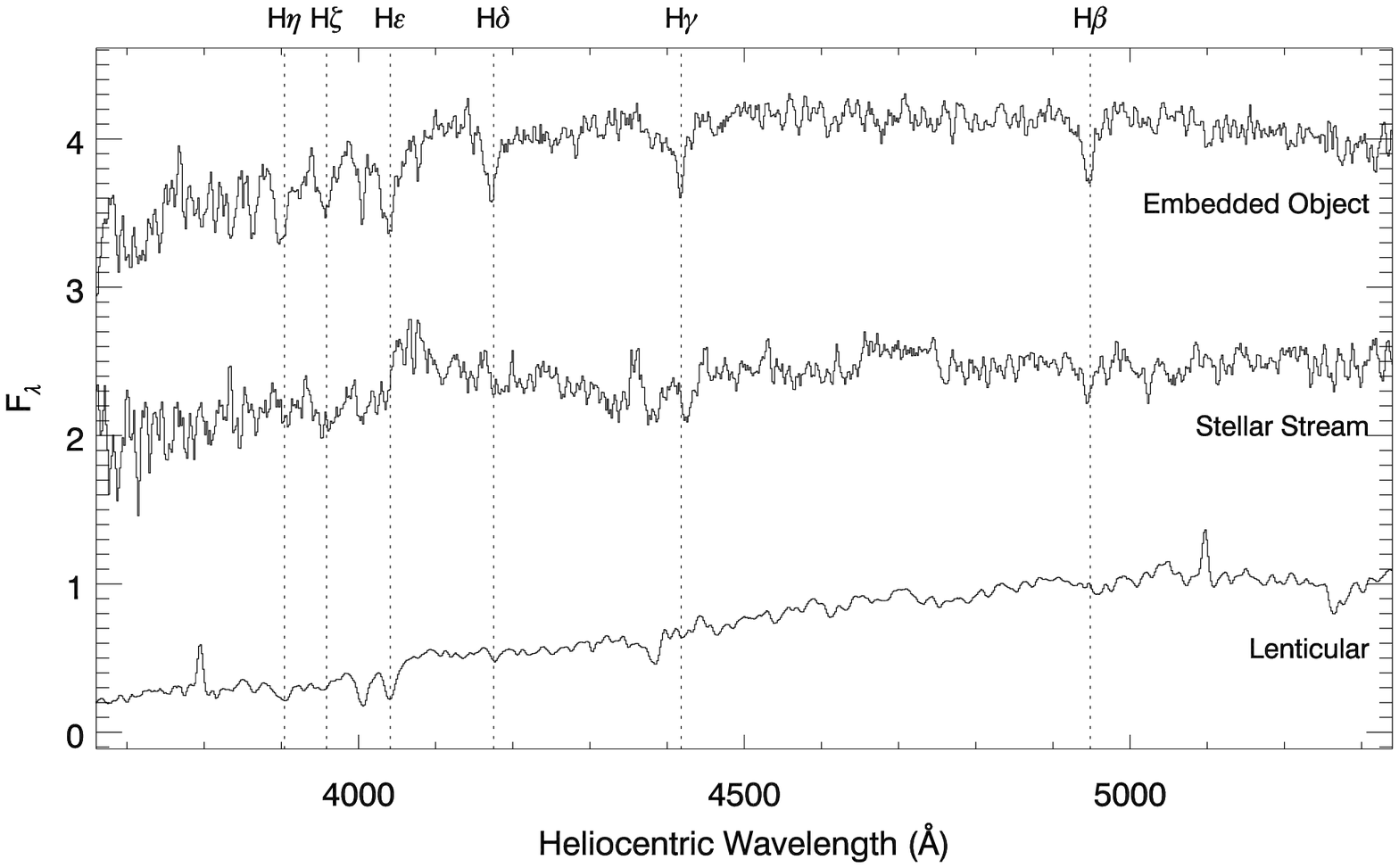}
\caption{The blue portion of our optical spectra of the lenticular region of \eso\ ({\em bottom}), the stellar stream of material ({\em middle}), and the compact object embedded within the stream ({\em top}).  The vertical axis is displayed in arbitrarily-scaled $F_{\lambda}$ units.  The dashed vertical lines indicate the positions of the H$\beta$-H$\eta$ absorption lines detected in the embedded source at $cz = 5370\pm50$~\kms.  The lenticular spectrum also contains [\ion{O}{2}] $\lambda$3717 and [\ion{O}{3}] $\lambda$5007 emission from a foreground \HII\ region in the slit (see Table~\ref{tab:lentlines}).
\label{fig:apospec}}
\end{figure}

The DIS spectrograph has a spatial scale of $0\farcs4$ per pixel, and extraction apertures of 10\arcsec, 40\arcsec, and 5\arcsec\ were used for the lenticular component of \eso, the stellar stream, and the source embedded in the stream, respectively. The spectra of these galaxy components are shown in Figure~\ref{fig:apospec}.  Only the blue portion of the spectra are shown, and the flux levels have been arbitrarily scaled for ease of display. The vertical dashed lines indicate the positions of the H$\beta$-H$\eta$ absorption lines detected in the spectrum of the embedded source.  There are indications of even higher-order Balmer absorption from \Ht, \Hi, and \Hk\ in the spectrum of the embedded source, but declining signal-to-noise blueward of 3800~\AA\ precludes us from making positive identifications.

Tables~\ref{tab:nuclines}-\ref{tab:lentlines} list the line identification, rest wavelength, observed wavelength, heliocentric velocity, and rest-frame equivalent width of all lines detected in the DIS spectra, including lines detected at wavelengths not displayed in Figure~\ref{fig:apospec}.  For the emission lines detected in the lenticular nucleus, Table~\ref{tab:lentlines} lists integrated line fluxes instead of rest-frame equivalent widths.  These fluxes have not been corrected for Galactic or intrinsic reddening.   Rest wavelengths for these and all other Tables were taken from the Atomic Line List\footnote{The Atomic Line List is hosted by the Department of Physics and Astronomy at the University of Kentucky (see \tt{http://www.pa.uky.edu/$\sim$peter/newpage/}).}.

Examination of these Tables shows that \CaII\ H \& K absorption lines are detected in all three spectra, that the embedded source is detected primarily in Balmer absorption lines, and that the lenticular component is detected mostly in nebular emission with some absorption from an older stellar population.  Further, the absorption lines detected in the three spectra indicate a clear hierarchy in the age of their stellar populations with the embedded source being the youngest, the stellar stream intermediate in age, and the lenticular nucleus the oldest \citep{bruzual03}.

There is also a significant velocity difference between the emission and absorption lines detected in the lenticular component of \eso, suggesting that the emission lines are associated with a foreground \HII\ region (see Table~\ref{tab:lentlines}).  The nebular emission-line velocity of this \HII\ region ($5380\pm10$~\kms) is similar to the peak \HI\ 21-cm emission velocity of region A (5338~\kms; see Table~\ref{tab:HI} and Fig.~\ref{fig:vlamap}), the nearest \HI\ 21-cm peak to the lenticular nucleus, confirming our earlier identification of the \HI\ 21-cm emission peaks A--C with spiral disk gas.  The absorption-line velocity in Table~\ref{tab:lentlines} ($5460\pm20$~\kms) then represents the velocity of the lenticular nucleus itself.

We have detected very weak \Ha\ emission from the \HII\ regions in the extended spiral arm at our third slit position.  These \HII\ regions, which are heavily blended at ground-based resolution, are located $\sim5\arcsec$ ($\sim\,$2\h~kpc) north of \pks\ (see Fig.~\ref{fig:hst}) and have a heliocentric velocity of $5440\pm20$~\kms.  Examination of the their 2D spectrum reveals a velocity gradient across the \Ha\ blend that we have extrapolated to the position of \pks\ to estimate the disk velocity at the quasar position to be $5500\pm30$~\kms.  This estimate is close to the \HI\ 21-cm emission velocity near the quasar position, which ranges from $\sim\,$5400-5490~\kms\ (CvG92).

\input tab2.tex
\input tab3.tex
\input tab4.tex

\section{Absorption Lines Associated with \eso\ in the Spectrum of \pks}
\label{qso}

Past studies of the optical and radio spectra of \pks\ have found absorption associated with the nearby galaxy \eso\ ($cz_{\rm gal} = 5380\pm10$~\kms).  A high-resolution ($v_{\rm res} \approx 30$~\kms) optical spectrum obtained by \citet{bergeron87} revealed two velocity components in the \NaI~D doublet at $5250\pm10$ and $5490\pm10$~\kms.  A lower resolution ($v_{\rm res} \approx 225$~\kms) optical spectrum, also from \citet{bergeron87}, showed \CaII~H \& K absorption at $5340\pm90$~\kms.  CvG92 also found two velocity components in the \HI\ 21-cm absorption spectrum of \pks, although their detection of the 5490~\kms\ component was tentative due to potential confusion from emission in the beam.  

We have acquired more sensitive \HI\ 21-cm and optical spectra of \pks, which we present in \S\,\ref{qso:vlaspec} and \S\,\ref{qso:apospec}, respectively.  We have also analyzed an archival \hst\ near-UV (NUV) spectrum of \pks, which is presented in \S\,\ref{qso:nuvspec} and Appendix~\ref{app:fos}.  These spectra allow us to confirm the presence of two \HI\ 21-cm absorbers, search for evidence of absorption from both velocity components in the \CaII~H \& K line profiles, and determine whether any gas associated with \eso\ is present in the NUV spectrum of \pks.

\subsection{\HI\ 21-cm Spectrum}
\label{qso:vlaspec}

\begin{figure}
\epsscale{1.00}
\centering \plotone{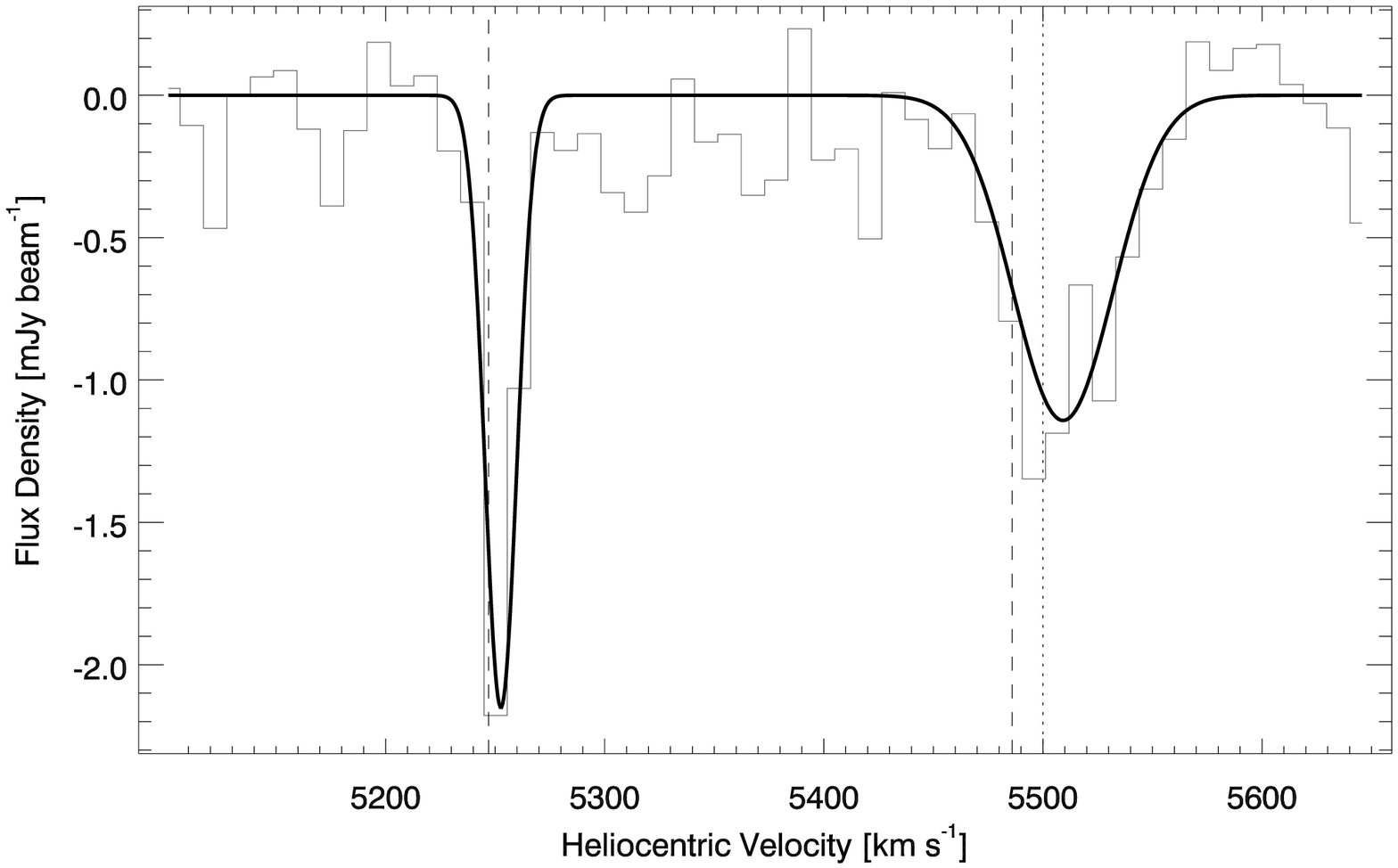}
\caption{The \HI\ 21-cm absorption spectrum of \pks\ at $20\arcsec \times 13\arcsec$ resolution and 10.4~\kms\,channel$^{-1}$ with best-fit absorption line profiles (see Table~\ref{tab:absfit}) overlaid.  The dashed vertical lines show the velocities of the \NaI~D absorption components found by \citet{bergeron87}.  The dotted vertical line indicates the disk velocity at the position of \pks, extrapolated from \HII\ regions in the extended spiral arm of \eso\ (see \S\,\ref{gal:apospec}).
\label{fig:vlaspec}}
\end{figure}

Our VLA observations (\S\,\ref{gal:vla}) allow us to search for \HI\ 21-cm absorption against the quasar continuum in addition to studying the \HI\ 21-cm emission properties of \eso.  The \HI\ 21-cm absorption spectrum of \pks\ is shown in Figure~\ref{fig:vlaspec} with Gaussian fits to the absorption lines overlaid with thick black lines.  The dashed vertical lines show the velocities of the \NaI~D absorption components found by \citet{bergeron87}, which clearly match the observed 21-cm absorption velocities.  Table~\ref{tab:absfit} lists best-fit physical parameters derived from the Gaussian fits to the \HI\ absorption components in Figure~\ref{fig:vlaspec}.

The \HII\ regions just NNW of \pks\ (see Fig.~\ref{fig:hst}) have heliocentric velocities comparable to the 5510~\kms\ \HI\ absorption complex.  An extrapolation of the \Ha\ emission line velocities towards \pks\  (see \S\,\ref{gal:apospec}) arrives at the velocity indicated by the dotted vertical line in Figure~\ref{fig:vlaspec}.  From this extrapolation we identify the 5510~\kms\ \HI\ 21-cm absorber with the extended spiral arm of \eso, and thus as galactic disk absorption. The velocity of the \HI\ 21-cm emission near the position of \pks\ is also compatible with this interpretation (CvG92).

CvG92 identified very weak \HI\ 21-cm emission aligned with the stellar stream of \eso\ and extending much further to the north (see their Figs.~3 and 4), but at significantly lower spatial resolution ($50\arcsec$) than we show in Figure~\ref{fig:vlamap}.  Regions D \& E in Figure~\ref{fig:vlamap} are the densest regions of this very diffuse emission, which extends into the vicinity of \pks\ at velocities between 5190 and 5296~\kms\ (CvG92).  CvG92 suggest that the \HI\ 21-cm absorber at 5255~\kms\ arises in this low surface brightness gas.  Given the complex optical and \HI\ 21-cm morphology of \eso, we hypothesize that the 5255~\kms\ absorber represents gas that has been tidally stripped from the disk of \eso.

The strength of \HI\ 21-cm absorption depends on the column density ($N_{\rm H\,I}$), covering fraction ($f$), and spin temperature ($T_{\rm spin}$) of the absorbing gas (see Table~\ref{tab:absfit}).  Assuming $T_{\rm spin} \sim 20$~K (\citealp{kobulnicky99} found $T_{\rm spin} = 20$--40~K in tidally disrupted gas in the Magellanic Bridge) and $f=1$, the column density of the 5255~\kms\ absorber is  $N_{\rm H\,I} \sim 5\times10^{18}~{\rm cm^{-2}}$ and the column density of the 5510~\kms\ absorber is $N_{\rm H\,I} \sim 9\times10^{18}~{\rm cm^{-2}}$.  These column densities are clearly uncertain due to ambiguities in both the spin temperature and covering factor of the absorbing gas; however, a higher spin temperature indicative of disk\footnote{The cold neutral medium of spiral galaxies typically has $T_{\rm spin} \sim 100$~K \citep[e.g.,][]{roy06}.} or halo\footnote{\citet{keeney05} found $T_{\rm spin} \sim 500$~K in the halo of the nearby spiral NGC~3067.} gas and/or a smaller covering factor would tend to \textbf{increase} the derived column density, so we treat the above estimates as lower limits to the true \HI\ column. Thus, the two 21-cm absorbers may combine to form a DLA ($N_{\rm H\,I} \geq 2\times10^{20}~{\rm cm^{-2}}$) as has been found in several other QSO sight lines that pass within 15\h~kpc of a low-$z$ foreground galaxy \citep[see, e.g., Table~2 of][]{zwaan05}.

The large velocity width of the 5510~\kms\ absorber could be caused by either thermal broadening or velocity structure along the line of sight.  The Gaussian shape of the line profile suggests that the broadening is thermal, but the associated kinetic temperature of the absorber is $\sim\,$60,000~K in that case.  The kinetic and spin temperatures should be coupled due to particle collisions; however, our kinetic temperature estimate for the 5510~\kms\ absorber is an order of magnitude greater than the largest spin temperatures reported in the literature \citep[$\sim\,$5,000~K; see, e.g.,][]{kanekar09}.  If we allow the kinetic and spin temperatures to be coupled at 60,000~K then the 5510~\kms\ absorber would have a neutral fraction of $\sim10^{-4}$ \citep{sutherland93} and an \HI\ column density of $N_{\rm H\,I} \gtrsim 10^{22}~{\rm cm^{-2}}$, or a total hydrogen column of $N_{\rm H} \sim 10^{26}~{\rm cm^{-2}}$.  This situation is clearly unphysical, since gas at these densities would be molecular and quickly cool to more modest temperatures \citep*[e.g.,][]{krumholz09}.  Therefore, we attribute the width of the 5510~\kms\ 21-cm absorber to velocity structure along the line of sight, which implies that the extended spiral arm of \eso\ has a velocity dispersion of $\sim\,$25~\kms\ at the quasar position.

\newpage
\subsection{Optical Quasar Spectrum}
\label{qso:apospec}

\citet{bergeron87} argued that the observed \CaII/\NaI\ ratio (\eqw(\CaII)/\eqw(\NaI$) = 0.34$ for the 5255 and 5510~\kms\ absorbers combined) points toward a galactic disk origin for the absorbing gas.  However, since their \CaII\ spectrum does not resolve both of the velocity components that they detected in \NaI, they cannot assess the origin of the individual absorbers directly.  Therefore, we obtained high-resolution optical spectra of \pks\ near the \CaII~H \& K ($v_{\rm res} \approx 150$~\kms) and \NaI~D ($v_{\rm res} \approx 75$~\kms) absorption lines associated with \eso\ to study the \CaII\,/ \NaI\ ratio of the individual velocity components.

\eso\ was observed with the DIS spectrograph of the ARC 3.5-m telescope at APO on 10~Feb~2010, 4~Apr~2010, and 6~May~2010 using a $1\farcs5$ slit and the B1200+R1200 gratings. At central wavelengths of 4500 and 6400~\AA, these gratings cover the wavelength ranges from 3900--5100~\AA\ and 5800--7000~\AA\ at 2~\AA\ resolution.  Observations of Feige~34 taken each night were used for flux calibration, and all data were reduced with a combination of standard IRAF tasks and custom IDL routines.

\begin{figure*}
\epsscale{1.00}
\centering \plotone{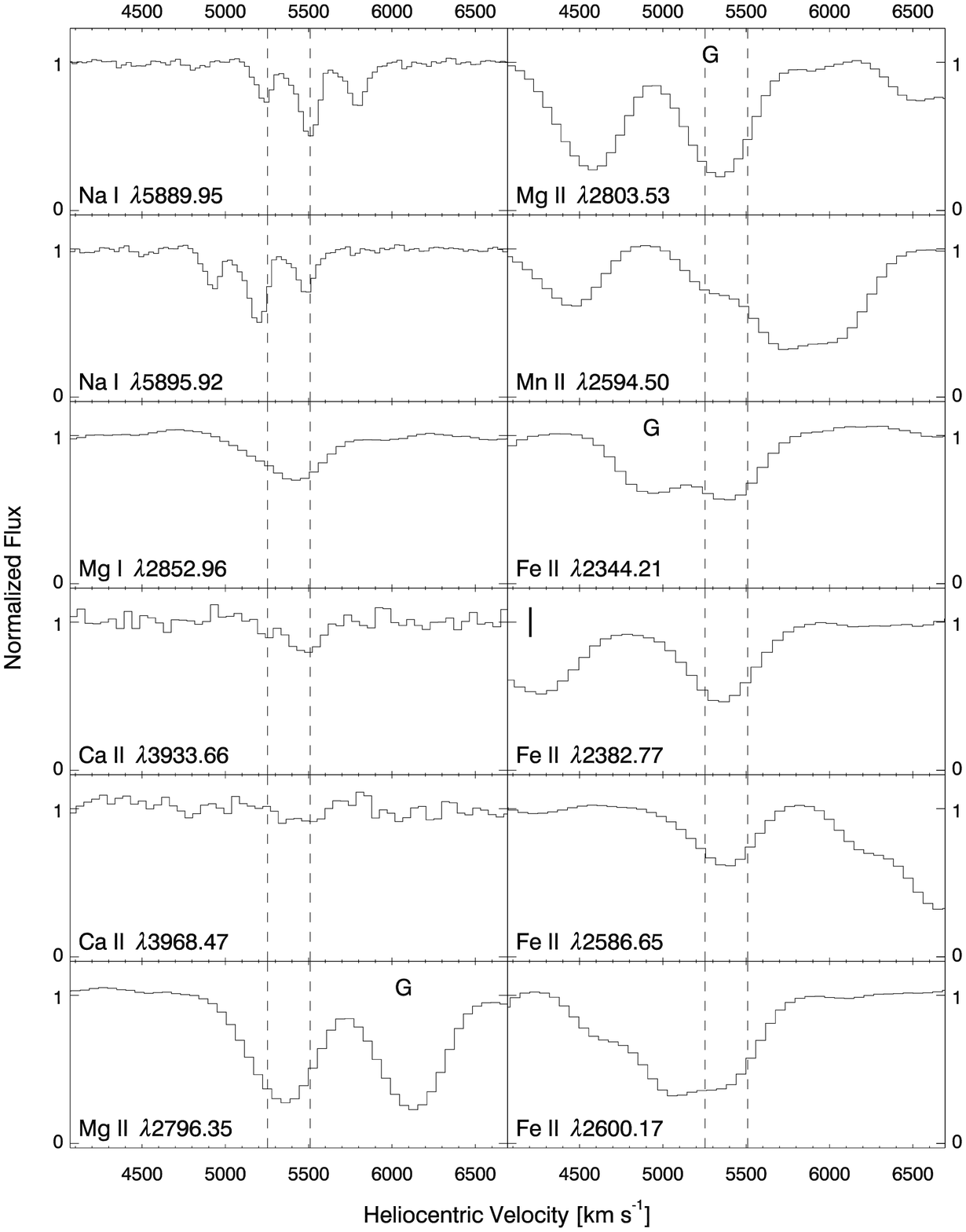}
\caption{Metal lines detected in the NUV and optical spectra of \pks\ at the redshift of \eso, arranged in order of increasing ionization potential.  The species and rest wavelength of the line are labelled in the lower-left corner of each panel.  The vertical dashed lines show the \HI\ 21-cm absorption velocities (see Table~\ref{tab:absfit}).  Positions of Galactic ISM absorption lines are marked with a ``G'' and lines associated with the LLS at $z=0.85238$ are marked with vertical tick marks.
\label{fig:pksspec}}
\end{figure*}

The regions surrounding the \CaII~H \& K and \NaI~D absorption lines associated with \eso\ are shown in Figure~\ref{fig:pksspec}.  Two velocity components are clearly detected in the \NaI~D doublet, with the \NaI~D1 line of the bluer component blended with the \NaI~D2 line of the redder component.  Best-fit velocities and deconvolved equivalent widths of these components were determined by simultaneous Voigt profile fits to the \NaI~D doublets.  The velocity structure of the \CaII\ profiles is less clear, but an apparent optical depth analysis of the \CaII~K profile shows two significant peaks, which we use to measure the velocities and calculate the apparent column densities of both components \citep{sembach92}.  Deconvolved equivalent widths were then calculated from the apparent column densities assuming optically thin absorption.  

Table~\ref{tab:CaNa} lists the line identification, rest wavelength, observed wavelength, heliocentric velocity, and rest-frame equivalent widths derived from these profiles.  The \CaII\ and \NaI\ velocities clearly agree with each other, as well as with the velocities of the \HI\ 21-cm components.  However, while the velocity of the apparent column density peak associated with the bluer \CaII\ component clearly matches that of the bluer \NaI\ component, its equivalent width is not significant and is hereafter treated as an upper limit.

The 5255 and 5510~\kms\ velocity components have \eqw(\CaII)/\eqw(\NaI) ratios of $<\,$0.26 and $0.63\pm0.15$, respectively, and a combined ratio of $<\,$0.48.  Ratios this low require a weak ionizing radiation field and are typically associated with disk gas in our Galaxy \citep{morton86}, whereas larger values (\eqw(\CaII)/\eqw(\NaI$) \ga 1$) indicate a stronger radiation field associated with a galactic halo environment \citep[e.g.,][]{morton86,stocke91}.  A disk gas origin for both velocity components suggests that the 5255~\kms\ absorber represents gas that has been tidally stripped from the disk of \eso\ rather than a high-velocity cloud in the halo of the galaxy.

\input tab5.tex

\subsection{\hst\ Near-UV Spectrum}
\label{qso:nuvspec}

\pks\ was observed with the Faint Object Spectrograph (FOS) on board \hst\ on 13~Jan~1997 for a total of 6540\,s with the G270H grating as part of GO program 5654.  These data were taken through the $1\farcs0$ aperture\footnote{The actual size of the aperture is $0\farcs86$ since these observations were obtained after the installation of COSTAR on \hst.} and cover a wavelength range of 2222-3277~\AA\ at 2~\AA\ ($\sim\,$200~\kms) resolution.  The FOS spectrum of \pks\ is filled with absorption lines from many redshifts, including Galactic ISM lines, lines associated with \eso, lines associated with the Lyman-limit system (LLS) at $z=0.85238$ \citep{bergeron87}, and intergalactic \lya\ forest absorbers.  We discuss the lines associated with \eso\ here and defer discussion of the other absorption lines in the \pks\ spectrum to Appendix~\ref{app:fos} since this spectrum has not been discussed previously in the literature.

Table~\ref{tab:foslines} lists the line identification, rest wavelength, observed wavelength, heliocentric velocity, and rest-frame equivalent width of all absorption lines associated with \eso\ that were detected at $>4\sigma$ confidence (observed $\eqw > 200$~m\AA) in the \hst/FOS spectrum of \pks\footnote{To ensure that Galactic ISM lines were centered at $v_{\rm helio} \sim 0$ we applied a zero-point offset of $145\pm8$~\kms\ to the FOS wavelength scale (see Appendix~\ref{app:fos} for more details).}.  The average velocty centroid of this absorption line system is $5360\pm10$~\kms, which is intermediate in velocity to the two \HI\ 21-cm components.  In addition to the lines listed in Table~\ref{tab:foslines}, the FOS spectrum shows indications of \MnII~$\lambda$2577 and \MnII~$\lambda$2606 absorption associated with \eso\ at 2-$3\sigma$ confidence.

\input tab6.tex

Figure~\ref{fig:pksspec} shows the absorption line profiles of all of the lines listed in Table~\ref{tab:foslines}.  Lines marked with a ``G'' are Galactic ISM lines and those marked with solid vertical lines are associated with the LLS at $z=0.85238$.  The dashed vertical lines show the velocities of the \HI\ 21-cm absorbers at 5255 and 5510~\kms.  Unfortunately, the velocity resolution of the FOS spectrum is comparable to the separation of the 21-cm velocity components.  Given that the average velocity centroid of the NUV line profiles is intermediate to the \HI\ 21-cm absorber velocities, it is likely that they represent blended absorption from both of the 21-cm velocity components.

\subsection{Summary of Absorption Lines Associated with \eso}
\label{qso:summary}

Two velocity components at 5255 and 5510~\kms\ have been detected in \HI\ 21-cm, \CaII, and \NaI\ absorption, and the low-resolution NUV absorption line profiles of \ion{Fe}{2}, \ion{Mn}{2}, \ion{Mg}{1}, and \ion{Mg}{2} likely represent blended absorption from both components.  The 5510~\kms\ absorber is coincident in velocity with disk gas in the extended spiral arm of \eso\ at the location of \pks\ as traced by \HI\ 21-cm emission (CvG92) and \HII\ regions in the arm (\S\,\ref{gal:apospec}).  The low \CaII\,/ \NaI\ ratio of this component also suggests a galactic disk origin for the absorbing gas.

The origins of the 5255~\kms\ component are less clear.  It has an unresolved \HI\ 21-cm absorption line profile and its \CaII/\NaI\ ratio is even lower than that found for the 5510~\kms\ component, both of which suggest that the 5255~\kms\ absorber is cold disk gas with a simple line-of-sight velocity structure.  However, the velocity of this component differs from the disk velocity at the quasar position by $\sim\,$250~\kms, which would usually imply a high-velocity cloud origin for the absorbing gas.  We believe that the 5255~\kms\ component represents gas that was tidally stripped from the western side of the inclined spiral disk of \eso\ as a result of a past merger and now resides in the very diffuse tidal tail detected in \HI\ 21-cm emission near the position of \pks\ (CvG92).  This origin simultaneously explains the disk gas signatures and high-velocity cloud kinematics of this velocity component.

One drawback to our \HI\ 21-cm, optical, and NUV spectra of \pks\ is that we are only sensitive to neutral or singly-ionized species at the redshift of \eso.  We are therefore unable to speculate on the multi-phase nature of the absorbing gas, which may be crucial for understanding the radiation field incident on the 5255 and 5510~\kms\ absorbers and thus their locations with respect to \eso.  Detailed study of the ionization state of the absorbing gas requires searching for Lyman series \HI\ lines or abundant metals of varying ionization states (e.g., \ion{C}{2}, \ion{C}{3}, \CIV, \ion{Si}{2}, \ion{Si}{3}, \ion{Si}{4}, \ion{N}{5}, \ion{O}{6}) in the far-UV (FUV) spectrum of \pks\ as has been done for other close QSO/galaxy pairs \citep[see, e.g.,][]{bowen01, bowen05, jenkins05, stocke04, keeney05, keeney06, chen09}.   

Unfortunately, there is a LLS at $z=0.85238$ that makes \pks\ too faint for FUV spectroscopy (see Appendix~\ref{app:fos} and \citealp{bergeron87}).  \pks\ was observed by the {\sl Galaxy Evolution Explorer} ({\sl GALEX}) as part of its All-Sky Imaging Survey and found to have a FUV magnitude of 21.2 ($F_{\lambda} \approx 1.5\times10^{-16}~{\rm ergs\,s^{-1}\,cm^{-2}\,\mbox{\AA}^{-1}}$).  However, even with the exquisite FUV sensitivity of the recently-installed Cosmic Origins Spectrograph on \hst\ \citep{green10,osterman10} it would still take $\sim\,$60 orbits to reach ${\rm S/N} = 10$ per resolution element ($\mathcal{R} \approx 18,000$) with the G160M grating ($\lambda \approx 1400$--1800~\AA), which covers the \ion{Si}{4} and \CIV\ transitions.

Two recent papers \citep{gupta10,borthakur10} have searched for \HI\ 21-cm absorption in quasar-galaxy pairs culled from the Sloan Digital Sky Survey (SDSS).  Both papers highlight the difficulty of detecting galaxies via \HI\ 21-cm absorption:  only 1 of 5 galaxies were detected by \citet{gupta10} and 1 of 12 by \citet{borthakur10}, and both of these galaxies had impact parameters of $\leq\,$11\h~kpc to the quasar sight line.  As for the \pks\ absorbers, \citet{gupta10} and \citet{borthakur10} found that their absorbers had DLA/sub-DLA column densities and were most likely associated with disk gas in the foreground galaxies.  The \citet{gupta10} absorber has a significantly larger \eqw(\CaII)/\eqw(\NaI) ratio ($1.19\pm0.19$) than our \pks\ absorbers, however, which suggests that it is embedded in a stronger radiation field.

\input tab7.tex

\section{Detection of An Ejected Galaxy Nucleus?}
\label{nucleus}

There are several potential identifications for the compact object embedded in the stellar stream of \eso.  The simplest explanation would be that it is a background elliptical galaxy projected in the middle of the stream by chance.  However, our optical spectrum of this source (Fig.~\ref{fig:apospec}) shows that its radial velocity is $5370\pm50$~\kms, $90\pm55$~\kms\ less than the absorption-line velocity of the lenticular nucleus to the south (see Tables~\ref{tab:nuclines} \& \ref{tab:lentlines}), so it is clearly not a background object.  

Another explanation would be that the compact object is a globular cluster.  Close examination of the embedded source in Figure~\ref{fig:hst} shows that it has a relatively bright core surrounded by a more extended spheroid of lower surface brightness material.  The core is marginally resolved with a FWHM of $\sim0\farcs2$ (70\h~pc) in our \hst\ images, which is $\sim\,$3 times larger than $\omega$~Cen, the largest globular cluster in the Milky~Way \citep*{tolstoy09}.  Furthermore, the core of the embedded object has an absolute B-band magnitude of $M_{\rm B} \approx -11.9$, corresponding to a luminosity of $L_{\rm B} \approx 10^7~L_{\Sun}$, which is $\sim\,$5 times larger than typical globular cluster values \citep[$M_{\rm B} \ga -10$;][]{tolstoy09}. The more extended emission has a FWHM of $\approx 2\arcsec$ (700\h~pc) and is $\sim\,$3.5 times brighter than the core of the embedded source.  This measurement is somewhat uncertain, however, due to the presence of the stellar stream in which it is embedded.

Tidal streams created by tidal disruption of a dwarf companion galaxy have been identified in the halo of our Galaxy \citep*{mathewson74,majewski03,newberg03,yanny03,rochapinto03} and in a few other nearby galaxies \citep[e.g., M~31;][]{ibata01,ferguson02,guhathakurta06,brown06}.  If both the compact source and the stellar stream of \eso\ were originally parts of a dwarf companion, it would have had $M_{\rm B} \la -16.0$ ($L_{\rm B} \ga 4\times10^8~L_{\Sun}$).  \citet*{grant05} found that $\ga\,$80\% of dwarf ellipticals in the Virgo cluster with $M_{\rm B} \leq -16$ had central nuclei.  Further, the two closest examples of nucleated dwarfs, the M~31 satellites M~32 and NGC~205, both show strong Balmer absorption in their nuclei \citep*{ho95} similar to the Balmer absorption seen in the compact object (see Fig.~\ref{fig:apospec}). Recent ground-based spectroscopy of dwarf ellipticals in the Fornax cluster and nearby galaxy groups \citep{koleva09} have also found evidence for gradients in age (increasing) and metallicity (decreasing) with increasing galaxy radius, which would explain the age differences between the compact object and the stellar stream (see \S\,\ref{gal:apospec}).  Given these compelling similarities we cannot be conclusive about ruling out a tidally-disrupted dwarf galaxy producing this stream and nucleus, but arguments against this hypothesis include: (1) This scenario requires a collision between three galaxies, not two, a much rarer occurrence; and (2) this explanation does not account for the absence of a second nucleus in the two galaxies to the south.  

Instead, we propose that the hyper-compact object in the stellar stream is the spiral's nucleus, ejected during the collision.  Given that the nucleus has an early-to-mid A star spectrum, as evidenced by strong Balmer absorption with no associated emission, an age for these stars is estimated to be $\sim\,$1~Gyr \citep{hansen94,bruzual03}. If this age indicates the time of a starburst created by the collision, then there is plenty of time for the nucleus to reach its current location 14\h~kpc in projection from the center of the spiral arms to the south and consistent with the (poorly constrained) radial velocity difference of $90\pm55$~\kms\ between the ejected nucleus and the lenticular nucleus.  The somewhat older stars in the stellar stream would constitute a small bulge of pre-existing stars also ejected during the collision, some with larger and some with smaller ejection velocities than the ejected nucleus.  However, the spiral would also have had time to move a comparable or larger distance on the sky since this putative merger-triggered starburst. So, by the ejected nucleus hypothesis, it appears likely that the current collision is not the first close passage of these two galaxies and we are seeing these galaxies during their second or third close passage.  

Alternatively, the age of the nuclear star cluster may not be related to the galaxy-galaxy collision, in which case the nucleus could have been ejected at a significantly higher velocity given the lack of proper motion between the lenticular and the spiral structure superposed on it; i.e., the collision which ejected the nucleus is very much still on-going.  The poor constraint on the velocity difference mentioned above and the unknown inclination angle of the ejected nucleus' path allow for a consistent picture but do not constrain this picture significantly.  So, the required scenario for an ejected spiral nucleus is somewhat complicated.  This unresolved discussion invites a numerical simulation to determine how easily these observables can be reproduced \citep[e.g.,][]{barnes09} and would lend considerable credence to the ejected nucleus hypothesis if they are reproducible.

Current simulations predict that there are two scenarios which can lead to the ejection of a galaxy nucleus as a result of the interaction of two galaxies:  grazing-incidence tidal interactions and mergers \citep[e.g.,][]{moore96,komossa08a,komossa08b}.  In both scenarios, the ejected nucleus is expected to consist of a supermassive black hole surrounded by a compact stellar system with a radius of a few tens of parsecs.  In the case of galaxy mergers, one of the nuclei can receive a kick of up to $\sim\,$1000~\kms\ from asymmetric gravitational wave emission associated with the merger, causing the displaced black hole and any stars within its gravitational sphere of influence to be ejected from the host galaxy \citep{komossa08a,komossa08b}.  Usually the so-called hypercompact galaxy nucleus that is ejected remains bound to the new system and eventually sinks via dynamical friction to its center where it will merge with the undisplaced nucleus.  The predictions of these simulations agree very well with what we observe in \eso.

While we do not know of a completely conclusive test of the ejected nucleus hypothesis, the detection of a weak AGN in this hyper-compact object would argue strongly in favor of that hypothesis.  Our VLA L-band continuum images set a 3$\sigma$ flux limit of $\la\,$1~mJy (corresponding to $L \la 10^{21}~{\rm W\,Hz^{-1}}$) at the position of this knot.  Since many elliptical galaxies have weak radio sources less luminous than this \citep{nagar02} this limit is not conclusive, especially in the context of an ejected spiral nucleus, which would normally be radio-quiet. 

\citet{loeb07} has assessed the circumstances by which a supermassive black hole and at least a portion of its accretion disk could be ejected from a galaxy merger in progress.  He finds that the disk accretion could survive for a few million years allowing the black hole to move $\sim\,$10~kpc away from the galaxy nucleus. Since this is the approximate distance that the candidate ejected nucleus identified here is from its putative nuclear location, there is some chance that it may be still accreting. A moderately deep {\sl Chandra} image might detect an X-ray source associated with a weak AGN in this nucleus and is likely the best observational test; i.e., an $\sim 10^7~M_{\Sun}$ black hole accreting at $\sim\,$1\% of the Eddington rate would be detectable with {\sl Chandra} in a few tens of ksecs.  

Recently, \citet{comerford09a,comerford09b} have proposed other candidate ejected nuclei based upon a large radial velocity difference between two emission line components in a distant AGN \citep[see also][]{boroson09}.  However, given the large cosmological distance to these ejected nuclei candidates, little detail can be discerned to confirm their hypothesis.

\section{Conclusions}
\label{conclusion}

In this paper we have presented new \hst\ imaging and supporting ground-based imaging and spectroscopy of the complex galaxy system \eso, which is located near on the sky to the background quasar \pks, whose optical, UV, and radio spectra contain absorption lines due to the foreground galaxy.  The \hst\ images reveal that a spiral galaxy is in the process of colliding with a lenticular galaxy since different nebular regions in the spiral arms are seen in front of or behind the lenticular's bulge.  Ground-based imaging and spectroscopy confirms that these nebular regions are \HII\ regions with velocities close to that of the lenticular's nucleus.  Thus, there is no doubt that these two systems are co-spatial (see Fig.~\ref{fig:hst}).  The rather diffuse \HI\ 21-cm emission associated with this system confirms this conclusion since, while some \HI\ is associated with the spiral arms in the late-type galaxy disk, much of it has been ejected from the spiral and is detected far to the north and east of the optical galaxy.  There may also be some diffuse \HI\ to the south of these galaxies. However, the size of the \HI\ emitting area of \eso\ is not appreciably larger than our estimates of the pre-merger size of the galaxy (see Fig.~\ref{fig:vlamap}).

The \hst\ and ground-based images also detect a ``stellar stream'' extending parallel to the \HI\ stream to the north, which we identify as a tidal feature also created in the collision.  Unique to this tidal feature (to our knowledge) the \hst\ images have discovered a marginally resolved source centered in the stream with an intermediate age ($\sim\,$1~Gyr) stellar population.  This knot is too luminous ($\sim 10^7~L_{\Sun}$) and too large to be a globular cluster and is unlikely to be the nucleus of a tidally-disrupted dwarf companion to these galaxies.  We suggest that this knot is the hypercompact galaxy nucleus of the inclined spiral component of \eso, which was ejected from the system as a result of its recent interaction with the lenticular component and is itself being shredded by tides to produce the observed stellar stream.  This assertion can be supported by detecting a non-thermal X-ray point source associated with the putative ejected nucleus and by numerical modeling of an early-type/late-type galaxy merger that results in the formation of a polar ring galaxy \citep{whitmore90,barnes09}.

The \pks\,/ \eso\ system provides us with a very detailed, and far from simple, view of a DLA/sub-DLA system.  Rather than an isolated gaseous disk of \HI, this DLA/sub-DLA is a mixture of two components, only one of which appears related to the spiral disk.  However, both have \CaII\,/ \NaI\ ratios typical of the disk of the Milky Way and not of halo gas.  In fact, the narrow high-velocity component has a smaller \CaII\,/ \NaI\ ratio than the broad \HI\ component we have identified as due to the outer spiral arm and so we also identify this as disk, not halo or HVC, material. Following CvG92, the best identification of the $\sim\,$5250~\kms\ component appears to be with the diffuse tidally-disrupted \HI\ found by those authors to the northeast of \eso\ at similar velocities.  

Using a spin temperature of 100~K, appropriate for galactic disk material, the column density associated with the outer spiral arm at $\sim\,$5500~\kms\ and the column density associated with the high-velocity component at $\sim\,$5250~\kms\ are nearly equal and, if added together, yield a total $\log{N_{\rm H\,I}} \approx 19.8~{\rm cm^{-2}}$, a sub-DLA.  However, using these estimated column densities and the velocity separation for the two components, we have simulated the \lya\ profile of this entire system and find it to be best-fit by a sub-DLA with a single velocity component having $\log{N_{\rm H\,I}} = 20.0$ when the signal-to-noise of the spectrum is modest ($\la\,$15).  This experiment suggests that using a single-component \lya\ profile to determine $N_{\rm H\,I}$ for DLAs and sub-DLAs with multiple metal-line components can overpredict the total \HI\ column density of the system, particularly for closely-spaced components with comparable $N_{\rm H\,I}$.

In summary, of the three low-$z$ QSO/galaxy pairs with likely DLA or sub-DLA absorbers that we have imaged using \hst, neither 3C~232\,/ NGC~3067 nor \pks\,/ \eso\ appear to have absorbers that are associated with quiescent thick disk gas.  This is in keeping with a recent suggestion by \citet{zwaan08} that the widths of low-ionization metal lines are too broad by a factor of two to be consistent with the canonical model of DLAs arising due to ordered rotation in cold disks \citep{wolfe86,wolfe95,lanzetta91}.  \citet{zwaan08} conclude that superwind outflows or galaxy interactions are most likely the cause of the broader velocity widths.  The current observational work supports this conclusion, particularly the case presented here.  

However, we do note that the third QSO/galaxy pair in our study, PKS~2020--370\,/ Klemola~31A does appear consistent with rotating disk gas.  CvG92 show that the observed \HI\ 21-cm emission contour at the quasar location is $\sim 2\times10^{20}~{\rm cm^{-2}}$ with velocities consistent with disk rotation.  Our \hst\ images show no \Ha\ emission beyond the spiral arms in Klemola~31A and neither line nor continuum emission is observed at the quasar location.  If we imagine that our small sample of \HI\ 21-cm and \CaII\,/ \NaI\ absorption-selected DLA/sub-DLAs are representative, then we expect that other DLA samples are a rather uniform mixture of disk gas, halo gas, and gas whose cross-section on the sky has been modestly increased by tidal interactions during galaxy collisions and mergers.

\acknowledgments B.\,A.\,K., J.\,T.\,S., and C.\,W.\,D. gratefully acknowledge support from NASA \hst\ General Observer grant GO-10925, as well as NASA grant NNX08AC14G. C.\,L.\,C. thanks the Max-Planck-Gesellschaft and the Humboldt- Stiftung for support through the Max-Planck-Forschungspreis.  We would also like to thank J.~Darling for many useful discussions as this work progressed.

This work is based on observations made with the NASA/ESA \hst, the Apache Point Observatory 3.5-m telescope, and the Very Large Array of the National Radio Astronomy Observatory (NRAO).  The \hst\ data were obtained at the Space Telescope Science Institute, which is operated by the Association of Universities for Research in Astronomy, Inc., under NASA contract NAS5-26555.  The Apache Point Observatory 3.5-m telescope is owned and operated by the Astrophysical Research Consortium.  The NRAO is a facility of the National Science Foundation operated under cooperative agremeent by Associated Universities, Inc.  This research has also made use of the NASA/IPAC Extragalactic Database, which is operated by the Jet Propulsion Laboratory, California Institute of Technology, under contract with the National Aeronautics and Space Administration.

\appendix
\section{Other Absorbers in the Near-UV Spectrum of \pks}
\label{app:fos}

\begin{figure}
\epsscale{1.00}
\centering \plotone{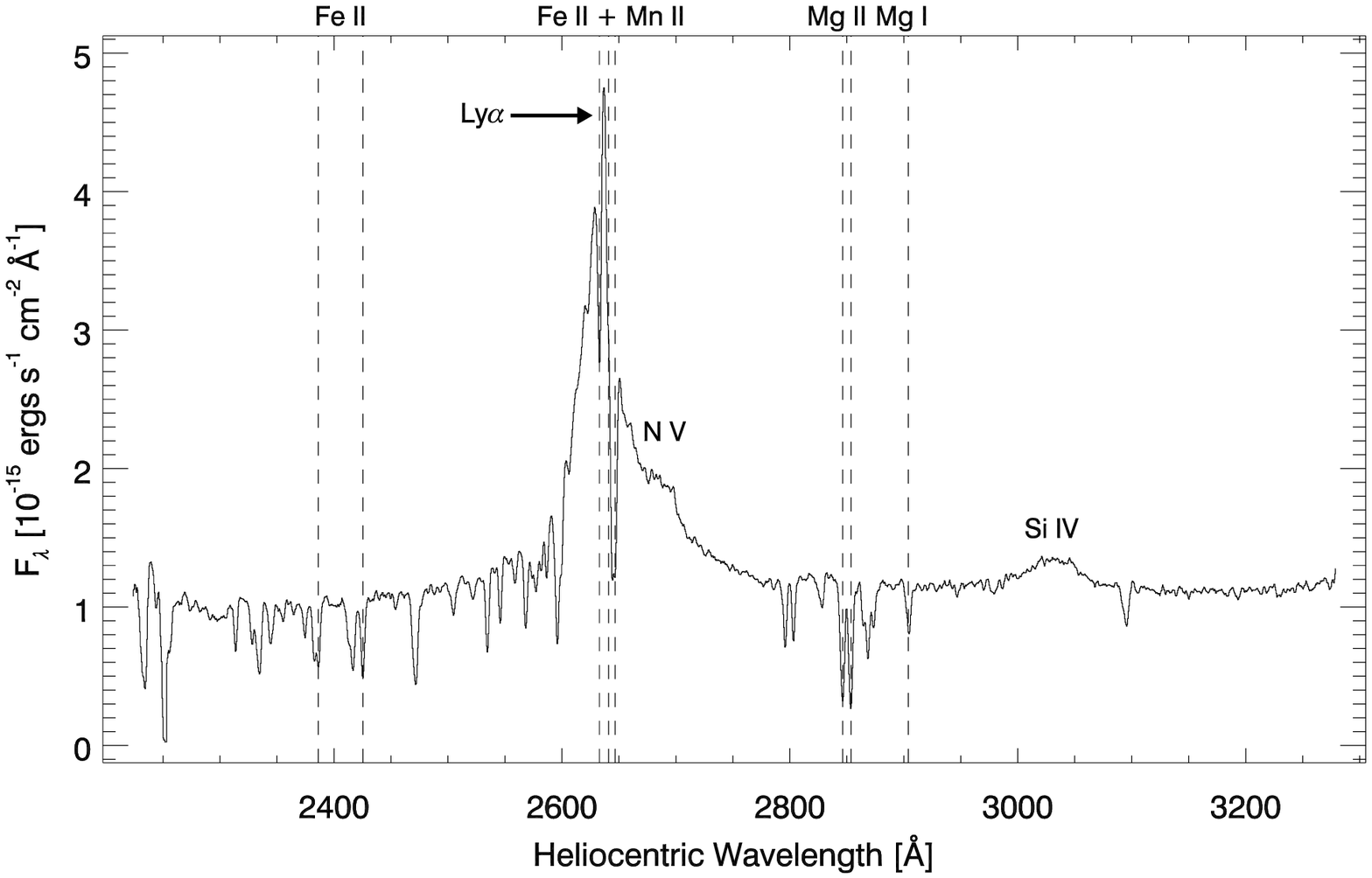}
\caption{The full \hst/FOS spectrum of \pks.  Quasar emission lines at $z=1.169$ are labelled and the positions of absorption lines associated with \eso\ (see \S\,\ref{qso:nuvspec} and Table~\ref{tab:foslines}) are indicated by the dashed vertical lines.
\label{fig:fosspec}}
\end{figure}

The \hst/FOS spectrum of \pks\ is filled with absorption lines from a variety of redshifts, including lines associated with \eso\ at $z=0.0178$ ($cz \approx 5500$~\kms; see \S\,\ref{qso:nuvspec}), lines associated with the LLS at $z=0.85238$ \citep{bergeron87}, intergalactic \lya\ forest absorbers, and Galactic ISM absorbers.  Here we present measurements of the absorption-line parameters of these systems.  Figure~\ref{fig:fosspec} displays the full FOS spectrum of \pks\ with quasar emission lines at $z=1.169$ labelled.  The dashed vertical lines indicate the positions of absorption lines associated with \eso\ (see \S\,\ref{qso:nuvspec}, Fig.~\ref{fig:pksspec}, and Table~\ref{tab:foslines}).

Table~\ref{tab:ISMlines} lists the line identification, rest wavelength, observed wavelength, heliocentric velocity, and rest-frame equivalent width of all Galactic ISM absorbers detected at $>4\sigma$ confidence.  We detect ISM absorption from four \ion{Fe}{2} transitions, two \MnII\ lines, the \MgII~$\lambda\lambda$2796, 2804 doublet, and \ion{Mg}{1}~$\lambda$2853.  We have added a zero-point offset of $145\pm8$~\kms\ to the FOS wavelength scale so that the interstellar absorption lines have an average velocity (excluding \MnII~$\lambda$2594) of $v_{\rm helio} = 0\pm10$~\kms.  This same zero-point offset has been added to all wavelengths, velocities, and redshifts in Tables~\ref{tab:foslines}--\ref{tab:IGMlines}.

\begin{figure}
\epsscale{1.00}
\centering \plotone{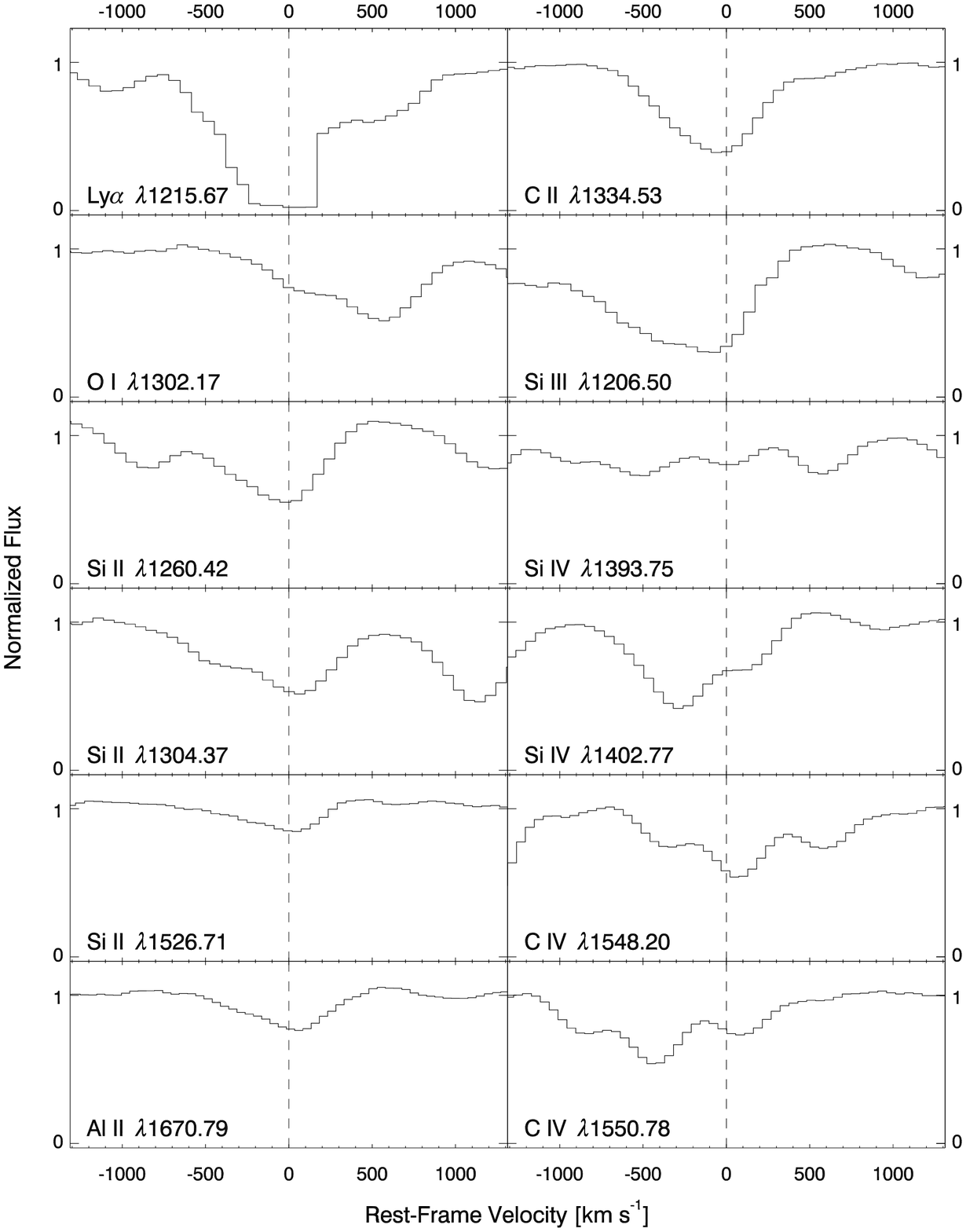}
\caption{Absorption lines associated with the LLS at $z=0.85238$ in the \hst/FOS spectrum of \pks, arranged in order of increasing ionization potential.  The species and rest wavelength of the line are labelled in the lower-left corner of each panel.  The vertical dashed line indicates a LLS rest-frame velocity of zero.
\label{fig:LLSlines}}
\end{figure}

The LLS is clearly detected in a number of neutral to moderately-ionized species.  Table~\ref{tab:LLSlines} lists the line identification, rest wavelength, observed wavelength, redshift, and rest-frame equivalent width of all lines detected at $>4\sigma$ confidence (observed $\eqw > 200$~m\AA) that are associated with the LLS.  The average redshift of these absorption lines is $z = 0.85238\pm0.00001$.  Figure~\ref{fig:LLSlines} shows the absorption line profiles of all lines listed in Table~\ref{tab:LLSlines}.

Finally, Table~\ref{tab:IGMlines} lists the line identification, rest wavelength, observed wavelength, redshift, and rest-frame equivalent width for all IGM absorbers detected at $>4\sigma$ confidence.  Fourteen \lya\ absorbers are detected at redshifts ranging from 0.84579 to 1.13536.  We also detect one \CIV~$\lambda$1548 absorber at $z=0.85021$ that lies $780\pm20$~\kms\ blueward of the \CIV~$\lambda\lambda$1548, 1551 absorbers associated with the LLS.  The \CIV~$\lambda$1551 absorption at $z=0.85021$ is blended with the \CIV~$\lambda$1548 absorption from the LLS, and the corresponding \lya\ absorber is blended with the LLS \lya.  While this absorber may represent a blue component to the LLS absorption, it is not present in any of the LLS absorption profiles in Figure~\ref{fig:LLSlines} except for \CIV.

\input tab8.tex
\input tab9.tex
\input tab10.tex


\end{document}

%% file: tab1.tex
\begin{deluxetable}{lcccc}

\tablecolumns{5}
\tablewidth{0pt}

\tablecaption{\HI\ 21-cm Emission Properties of \eso\
\label{tab:HI}}

\tablehead{\colhead{Region\tablenotemark{a}} & \colhead{RA} & \colhead{Dec} & \colhead{$v_{\rm peak}$\tablenotemark{b}} & \colhead{$M_{\rm H\,I}$} \\ & (J2000.0) & (J2000.0) & \colhead{(\kms)} & \colhead{($10^8\,h_{70}^{-2}~M_{\Sun}$)}}

\startdata
A & $13~30~05.2$ & $-20~55~56$ & 5338 & $2.61\pm0.33$ \\
B & $13~30~06.7$ & $-20~55~49$ & 5402 & $1.81\pm0.23$ \\
C & $13~30~06.4$ & $-20~56~18$ & 5306 & $1.91\pm0.25$ \\
D & $13~30~07.7$ & $-20~55~18$ & 5263 & $1.01\pm0.13$ \\
E & $13~30~07.6$ & $-20~54~38$ & 5253 & $2.22\pm0.28$
\enddata

\tablenotetext{a}{Region of interest as indicated in Fig.~\ref{fig:vlamap}.}
\tablenotetext{b}{The channel velocity of the \HI\ 21-cm emission peak for this region.}

\end{deluxetable}

%% file: tab2.tex
\begin{deluxetable}{lcccc}

\tablecolumns{5}
\tablewidth{0pt}

\tablecaption{Absorption Lines Detected in the Embedded Source
\label{tab:nuclines}}

\tablehead{\colhead{ID} & \colhead{$\lambda_{\rm rest}$} & \colhead{$\lambda_{\rm obs}$} & \colhead{$v_{\rm helio}$} & \colhead{\eqw} \\ & \colhead{(\AA)} & \colhead{(\AA)} & \colhead{(\kms)} & \colhead{(\AA)}}

\startdata
                    \Heta & 3835.38 & $3900.2\pm2.4$ & $5090\pm190$ & $13.7\pm9.7$ \\
                      \Hz & 3889.05 & $3957.4\pm3.7$ & $5290\pm290$ & $\phn7.8\pm6.7$ \\
                 Ca\,II~K & 3933.66 & $4005.2\pm1.4$ & $5480\pm110$ & $\phn3.3\pm2.3$ \\
     \He\tablenotemark{a} & 3970.07 & $4040.1\pm3.9$ & $5310\pm300$ & $15.4\pm8.9$ \\
                      \Hd & 4101.73 & $4172.5\pm2.6$ & $5190\pm190$ & $\phn9.0\pm2.7$ \\
                      \Hg & 4340.46 & $4416.7\pm2.9$ & $5290\pm200$ & $\phn9.3\pm3.5$ \\
                      \Hb & 4861.32 & $4946.4\pm2.2$ & $5270\pm140$ & $\phn7.4\pm2.4$ \\
                  Na\,I~D & 5891.94 & $5995.6\pm4.8$ & $5300\pm250$ & $\phn3.2\pm1.5$ \\
                      \Ha & 6562.80 & $6683.2\pm2.3$ & $5520\pm110$ & $\phn3.7\pm1.0$
\enddata

\tablecomments{The average absorption-line velocity for this source is $5370\pm50$~\kms.}
\tablenotetext{a}{This line is blended with Ca\,II~H but we attribute most of the absorption to \He\ given the presence of higher-order Balmer absorption in this spectrum.}

\end{deluxetable}

%% file: tab3.tex
\begin{deluxetable}{lcccc}

\tablecolumns{5}
\tablewidth{0pt}

\tablecaption{Absorption Lines Detected in the Stellar Stream
\label{tab:streamlines}}

\tablehead{\colhead{ID} & \colhead{$\lambda_{\rm rest}$} & \colhead{$\lambda_{\rm obs}$} & \colhead{$v_{\rm helio}$} & \colhead{\eqw} \\ & \colhead{(\AA)} & \colhead{(\AA)} & \colhead{(\kms)} & \colhead{(\AA)}}

\startdata
  Ca\,II~K & 3933.66 & $4006.8\pm3.9$ & $5600\pm300$ & $3.8\pm2.4$ \\
  Ca\,II~H & 3968.47 & $4036.7\pm2.6$ & $5180\pm190$ & $3.3\pm1.7$ \\
    G-band & 4304.40 & $4382.4\pm2.2$ & $5460\pm160$ & $4.7\pm3.6$ \\
       \Hg & 4340.46 & $4425.7\pm1.9$ & $5910\pm130$ & $5.9\pm3.7$ \\
       \Hb & 4861.32 & $4945.9\pm5.2$ & $5240\pm320$ & $4.8\pm2.4$
\enddata

\tablecomments{The average absorption-line velocity for this source is $5590\pm80$~\kms.}

\end{deluxetable}

%% file: tab4.tex
\begin{deluxetable*}{lccccclcccc}

\tabletypesize{\footnotesize}

\tablecolumns{11}
\tablewidth{0pt}

\tablecaption{Lines Detected in the Lenticular Nucleus
\label{tab:lentlines}}

\tablehead{\multicolumn{5}{c}{Emission Lines} & & \multicolumn{5}{c}{Absorption Lines} \\ \cline{1-5} \cline{7-11} \\ \colhead{ID} & \colhead{$\lambda_{\rm rest}$} & \colhead{$\lambda_{\rm obs}$} & \colhead{$v_{\rm helio}$} & \colhead{Integ. Flux} & \phantom{ID} & \colhead{ID} & \colhead{$\lambda_{\rm rest}$} & \colhead{$\lambda_{\rm obs}$} & \colhead{$v_{\rm helio}$} & \colhead{\eqw} \\ & \colhead{(\AA)} & \colhead{(\AA)} & \colhead{(\kms)} & \colhead{($10^{-16}~{\rm ergs\,s^{-1}\,cm^{-2}}$)} & & & \colhead{(\AA)} & \colhead{(\AA)} & \colhead{(\kms)} & \colhead{(\AA)}}

\startdata
 ${\rm [O\,II]}$ & 3727.42 & $3794.7\pm0.5$ & $5400\pm40$ & $\phn9.4\pm3.6$ & &         Ca\,II~K & 3933.66 & $4005.9\pm0.6$ & $5500\pm50$ & $10.3\pm3.1$ \\
${\rm [O\,III]}$ & 5007.84 & $5096.9\pm0.4$ & $5330\pm30$ & $10.4\pm2.4$    & &         Ca\,II~H & 3968.47 & $4041.1\pm0.7$ & $5480\pm50$ & $10.8\pm3.0$ \\
 ${\rm [N\,II]}$ & 6548.04 & $6664.2\pm1.2$ & $5310\pm60$ & $\phn3.3\pm0.8$ & &           G-band & 4304.40 & $4382.4\pm1.0$ & $5430\pm70$ & $\phn6.6\pm2.4$ \\
             \Ha & 6562.80 & $6680.6\pm0.3$ & $5380\pm20$ & $15.1\pm1.6$    & &          Na\,I~D & 5891.94 & $5999.2\pm0.6$ & $5450\pm30$ & $\phn4.9\pm0.7$ \\
 ${\rm [N\,II]}$ & 6583.46 & $6701.9\pm0.3$ & $5390\pm20$ & $14.6\pm1.6$ \\
 ${\rm [S\,II]}$ & 6716.44 & $6837.3\pm1.0$ & $5390\pm40$ & $\phn5.7\pm1.0$ \\
 ${\rm [S\,II]}$ & 6730.82 & $6852.0\pm0.8$ & $5390\pm40$ & $\phn6.1\pm1.0$
\enddata

\tablecomments{The average emission-line velocity for this source is $5380\pm10$~\kms and the average absorption-line velocity is $5460\pm20$~\kms.}

\end{deluxetable*}

%% file: tab5.tex
\begin{deluxetable}{cccc}

\tablecolumns{4}
\tablewidth{0pt}

\tablecaption{Gaussian Fits to \HI\ 21~cm Absorbers Detected Toward \pks
\label{tab:absfit}}

\tablehead{\colhead{$v_{\rm helio}$} & \colhead{FWHM} & \colhead{$\tau_0$} & \colhead{$N_{\rm H\,I} / (T_{\rm spin}/f)$} \\ \colhead{(\kms)} & \colhead{(\kms)} & & \colhead{($10^{17}~{\rm cm^{-2}\,K^{-1}}$)}}

\startdata
$5255\pm\phn5$ & $17\pm\phn7$ & $0.0068\pm0.0006$ & $2.3\pm1.0$ \\
$5510\pm10$    & $54\pm18$    & $0.0042\pm0.0006$ & $4.3\pm1.6$
\enddata

\end{deluxetable}

%% file: tab6.tex
\begin{deluxetable}{lcccc}

\tablecolumns{5}
\tablewidth{0pt}
\tablecaption{Lines Associated with \eso\ in the Optical Spectrum of \pks
\label{tab:CaNa}}

\tablehead{\colhead{ID} & \colhead{$\lambda_{\rm rest}$} & \colhead{$\lambda_{\rm obs}$} & \colhead{$v_{\rm helio}$} & \colhead{\eqw} \\ & \colhead{(\AA)} & \colhead{(\AA)} & \colhead{(\kms)} & \colhead{(m\AA)}}

\startdata
\CaII~K  & 3933.663 & $4002.7\pm0.8$ & $5260\pm60$    & $170\pm140$ \\
         &          & $4005.5\pm0.5$ & $5470\pm40$    & $590\pm190$ \\
\CaII~H  & 3968.469 & $4038.1\pm0.8$ & $5260\pm60$    & $\phn90\pm\phn80$\tablenotemark{a} \\
         &          & $4040.9\pm0.5$ & $5470\pm40$    & $330\pm110$\tablenotemark{a} \\
%
\NaI~D2  & 5889.950 & $5992.9\pm0.2$ & $5240\pm10$    & $610\pm\phn80$ \\
         &          & $5997.7\pm0.2$ & $5480\pm10$    & $830\pm\phn90$ \\
\NaI~D1  & 5895.924 & $5998.9\pm0.2$ & $5240\pm10$    & $400\pm\phn70$ \\
         &          & $6003.8\pm0.2$ & $5480\pm10$    & $640\pm\phn80$
\enddata

\tablenotetext{a}{The equivalent widths for the \CaII~H lines were calculated from the apparent column densities measured in the \CaII~K line profiles, assuming optically thin absorption.}

\end{deluxetable}

%% file: tab7.tex
\begin{deluxetable}{lcccc}

\tablecolumns{5}
\tablewidth{0pt}
\tablecaption{Lines Associated with \eso\ in the \hst/FOS Spectrum of \pks
\label{tab:foslines}}

\tablehead{\colhead{ID} & \colhead{$\lambda_{\rm rest}$} & \colhead{$\lambda_{\rm obs}$} & \colhead{$v_{\rm helio}$} & \colhead{\eqw} \\ & \colhead{(\AA)} & \colhead{(\AA)} & \colhead{(\kms)} & \colhead{(m\AA)}}

\startdata
                   Fe\,II & 2344.21 & $2386.4\pm0.2$ & $5390\pm30$ & $1400\pm170$ \\
                   Fe\,II & 2382.77 & $2425.3\pm0.2$ & $5350\pm20$ & $2140\pm100$ \\
                   Fe\,II & 2586.65 & $2632.9\pm0.2$ & $5360\pm20$ & $1360\pm140$ \\
                   Mn\,II & 2594.50 & $2641.4\pm0.2$ & $5420\pm20$ & $1590\pm180$ \\
                   Fe\,II & 2600.17 & $2645.7\pm0.2$ & $5250\pm20$ & $3440\pm160$ \\
                   Mg\,II & 2796.35 & $2846.2\pm0.2$ & $5350\pm20$ & $3180\pm160$ \\
  Mg\,II\tablenotemark{a} & 2803.53 & $2853.4\pm0.2$ & $5340\pm20$ & $3290\pm150$ \\
                    Mg\,I & 2852.96 & $2904.4\pm0.2$ & $5400\pm20$ & $1290\pm120$
\enddata

\tablecomments{The average velocity of this absorption-line system is $5360\pm10$~\kms.}
\tablenotetext{a}{This line is blended with Mg\,I~$\lambda$2853 absorption from the Galactic ISM.}

\end{deluxetable}

%% file: tab8.tex
\begin{deluxetable}{lcccc}

\tablecolumns{5}
\tablewidth{0pt}
\tablecaption{Galactic ISM Lines in the \hst/FOS Spectrum of \pks
\label{tab:ISMlines}}

\tablehead{\colhead{ID} & \colhead{$\lambda_{\rm rest}$} & \colhead{$\lambda_{\rm obs}$} & \colhead{$v_{\rm helio}$} & \colhead{\eqw} \\ & \colhead{(\AA)} & \colhead{(\AA)} & \colhead{(\kms)} & \colhead{(m\AA)}}

\startdata
                   Fe\,II & 2344.21 & $2344.6\pm 0.2$ & $\phantom{-}50\pm30$    & $1040\pm330$ \\
                   Fe\,II & 2374.46 & $2374.4\pm 0.2$ & $\phantom{-}\phn0\pm30$ & $\phn570\pm130$ \\
                   Fe\,II & 2382.77 & $2382.7\pm 0.2$ & $-10\pm30$              & $1310\pm150$ \\
                   Mn\,II & 2576.88 & $2577.3\pm 0.3$ & $\phantom{-}50\pm30$    & $1100\pm340$ \\
                   Fe\,II & 2586.65 & $2586.7\pm 0.2$ & $\phantom{-}10\pm20$    & $\phn840\pm270$ \\
  Mn\,II\tablenotemark{a} & 2594.50 & $2595.9\pm 0.2$ & $\phn160\pm20$              & $2460\pm180$ \\
                   Mg\,II & 2796.35 & $2795.6\pm 0.2$ & $-80\pm20$              & $1500\pm170$ \\
                   Mg\,II & 2803.53 & $2803.2\pm 0.2$ & $-30\pm20$              & $1170\pm150$ \\
   Mg\,I\tablenotemark{b} & 2852.96 & $2853.4\pm 0.2$ & $\phantom{-}50\pm20$    & $3350\pm150$
\enddata

\tablenotetext{a}{This line is blended with an IGM \lya\ absorber at $z=1.13536$.}
\tablenotetext{b}{This line is blended with Mg\,II~$\lambda$2804 absorption from \eso.}

\end{deluxetable}

%% file: tab9.tex
\begin{deluxetable}{lcccc}

\tablecolumns{5}
\tablewidth{0pt}
\tablecaption{Lines Associated with the LLS in the \hst/FOS Spectrum of \pks
\label{tab:LLSlines}}

\tablehead{\colhead{ID} & \colhead{$\lambda_{\rm rest}$} & \colhead{$\lambda_{\rm obs}$} & \colhead{$z$} & \colhead{\eqw} \\ & \colhead{(\AA)} & \colhead{(\AA)} & & \colhead{(m\AA)}}

\startdata
                  Si\,III & 1206.50 & $2233.3\pm0.1$ & $0.85104\pm0.00004$ & $1960\pm190$ \\
                     \lya & 1215.67 & $2251.5\pm0.1$ & $0.85203\pm0.00004$ & $3270\pm400$ \\
                   Si\,II & 1260.42 & $2334.2\pm0.1$ & $0.85193\pm0.00004$ & $1400\pm320$ \\
                     O\,I & 1302.17 & $2412.8\pm0.1$ & $0.85292\pm0.00005$ & $\phn580\pm\phn50$ \\
  Si\,II\tablenotemark{a} & 1304.37 & $2416.8\pm0.1$ & $0.85286\pm0.00004$ & $1060\pm\phn50$ \\
                    C\,II & 1334.53 & $2471.4\pm0.1$ & $0.85187\pm0.00004$ & $1610\pm\phn90$ \\
                   Si\,IV & 1393.75 & $2581.9\pm0.2$ & $0.85249\pm0.00006$ & $\phn320\pm140$ \\
                   Si\,IV & 1402.77 & $2599.7\pm0.1$ & $0.85323\pm0.00004$ & $\phn390\pm\phn70$ \\
                   Si\,II & 1526.71 & $2827.7\pm0.1$ & $0.85216\pm0.00005$ & $\phn380\pm\phn90$ \\
                    C\,IV & 1548.20 & $2868.5\pm0.1$ & $0.85281\pm0.00004$ & $\phn910\pm\phn80$ \\
                    C\,IV & 1550.78 & $2873.5\pm0.1$ & $0.85293\pm0.00004$ & $\phn550\pm\phn80$ \\
                   Al\,II & 1670.79 & $3094.9\pm0.1$ & $0.85236\pm0.00004$ & $\phn630\pm\phn70$
\enddata

\tablecomments{The average redshift of this absorption-line system is $0.85238\pm0.00001$.}
\tablenotetext{a}{This line is blended with Fe\,II~$\lambda$2374 absorption from \eso\ but we attribute most of the absorption to Si\,II at $z=0.85286$.}

\end{deluxetable}

%% file: tab10.tex
\begin{deluxetable}{lcccc}

\tablecolumns{5}
\tablewidth{0pt}
\tablecaption{IGM Lines in the \hst/FOS Spectrum of \pks
\label{tab:IGMlines}}

\tablehead{\colhead{ID} & \colhead{$\lambda_{\rm rest}$} & \colhead{$\lambda_{\rm obs}$} & \colhead{$z$} & \colhead{\eqw} \\ & \colhead{(\AA)} & \colhead{(\AA)} & & \colhead{(m\AA)}}

\startdata
                     \lya & 1215.67 & $2243.9\pm0.2$ & $0.84579\pm0.00007$ & $\phn240\pm140$ \\
                     \lya & 1215.67 & $2256.7\pm0.1$ & $0.85637\pm0.00005$ & $\phn460\pm160$ \\
                     \lya & 1215.67 & $2313.7\pm0.1$ & $0.90322\pm0.00005$ & $\phn420\pm110$ \\
                     \lya & 1215.67 & $2328.6\pm0.1$ & $0.91547\pm0.00005$ & $\phn360\pm130$ \\
                     \lya & 1215.67 & $2355.1\pm0.3$ & $0.93728\pm0.00012$ & $\phn180\pm100$ \\
                     \lya & 1215.67 & $2454.1\pm0.2$ & $1.01875\pm0.00006$ & $\phn170\pm\phn60$ \\
                     \lya & 1215.67 & $2504.7\pm0.1$ & $1.06037\pm0.00005$ & $\phn370\pm\phn80$ \\
                     \lya & 1215.67 & $2521.9\pm0.1$ & $1.07452\pm0.00006$ & $\phn230\pm\phn70$ \\
                     \lya & 1215.67 & $2534.5\pm0.1$ & $1.08490\pm0.00004$ & $\phn720\pm\phn60$ \\
                     \lya & 1215.67 & $2545.9\pm0.1$ & $1.09421\pm0.00004$ & $\phn420\pm\phn50$ \\
                     \lya & 1215.67 & $2558.9\pm0.1$ & $1.10492\pm0.00005$ & $\phn180\pm100$ \\
                     \lya & 1215.67 & $2568.3\pm0.1$ & $1.11270\pm0.00004$ & $\phn730\pm150$ \\
                     \lya & 1215.67 & $2573.5\pm0.2$ & $1.11697\pm0.00008$ & $\phn230\pm120$ \\
    \lya\tablenotemark{a} & 1215.67 & $2595.9\pm0.1$ & $1.13536\pm0.00004$ & $1150\pm\phn80$ \\
                    C\,IV & 1548.20 & $2864.5\pm0.1$ & $0.85021\pm0.00004$ & $\phn430\pm\phn70$
\enddata

\tablenotetext{a}{This line is blended with Mn\,II~$\lambda$2594 absorption from the Galactic ISM.}

\end{deluxetable}